\documentclass[10pt]{article}

\usepackage{amsmath}
\usepackage{amssymb}

\usepackage{graphicx}

\usepackage{cite}

\usepackage{color}

\usepackage{afterpage}

\usepackage{lscape} 

\newcommand{\rev}[1]{#1}

\usepackage[normalem]{ulem}

\usepackage[labelfont=bf,labelsep=period,justification=raggedright]{caption}

\makeatletter
\renewcommand{\@biblabel}[1]{\quad#1.}
\makeatother

\date{}

\begin{document}

\begin{flushleft}{\Large \textbf{
            Joint scaling laws in functional and evolutionary categories in
      prokaryotic genomes.\\  }
}

J.~Grilli$^{1}$,
B.~Bassetti$^{1,2}$,
Sergei~Maslov$^{3}$,
M.~Cosentino~Lagomarsino$^{4,5,\ast}$
\\
\bf{1} Dipartimento di Fisica, Universit\`a degli Studi di Milano, Milano, Italy
\\
\bf{2} I.N.F.N. Milano, Milano, Italy
\\
\bf{3} Department of Condensed Matter Physics and Materials Science,
Brookhaven National Laboratory, Upton, NY 11973, United States of America
\\
\bf{4} G\'enophysique / Genomic Physics Group, UMR 7238 CNRS
  ``Microorganism Genomics'', Paris, France
\\
\bf{5} Universit\'e Pierre et Marie Curie, Paris, France
\\
$\ast$ E-mail: Marco.Cosentino-Lagomarsino@upmc.fr
\end{flushleft}

\section*{Abstract}

  We propose and study a \rev{class-expansion}/innovation/loss model
  of genome evolution taking into account biological roles of genes
  and their constituent domains. In our model numbers of genes in
  different functional categories are coupled to each other.  For
  example, an increase in the number of metabolic enzymes in a genome
  is usually accompanied by addition of new transcription factors
  regulating these enzymes. Such coupling can be thought of as a
  proportional ``recipe'' for genome composition of the type ``a
  spoonful of sugar for each egg yolk''.  The model jointly reproduces
  two known empirical laws: the distribution of family sizes and the
  nonlinear scaling of the number of genes in certain functional
  categories (e.g. transcription factors) with genome size.  In
  addition, it allows us to derive a novel relation between the
  exponents characterizing these two scaling laws, establishing a
  direct quantitative connection between evolutionary and functional
  categories.  It predicts that functional categories that grow
  faster-than-linearly with genome size to be characterized by
  flatter-than-average family size distributions. This relation is
  confirmed by our bioinformatics analysis of prokaryotic
  genomes. This proves that the joint quantitative trends of
  functional and evolutionary classes can be understood in terms of
  evolutionary growth with proportional recipes.


\section{Introduction}

\begin{figure}[t]
\begin{center}
\includegraphics[width=0.9\textwidth]{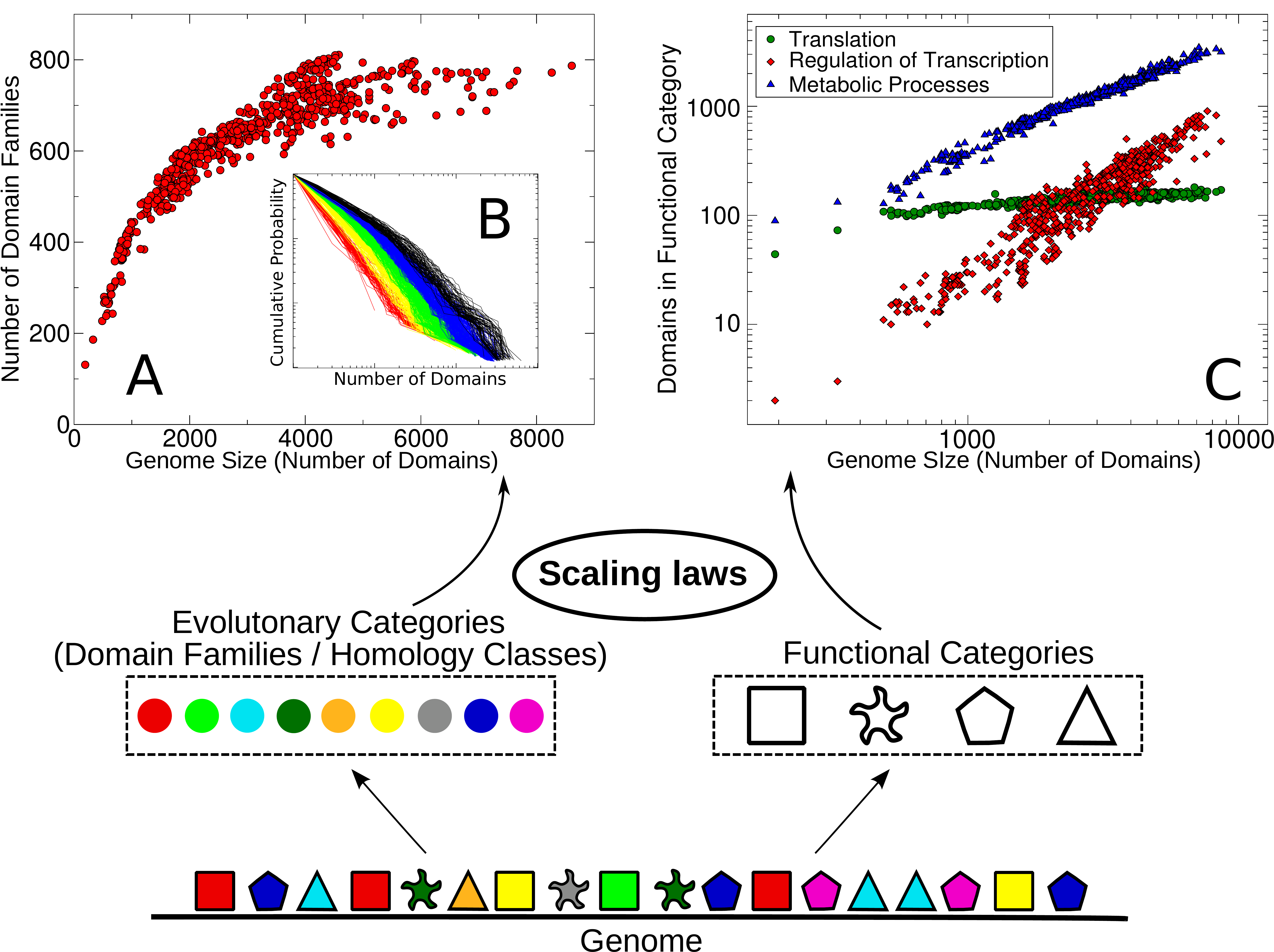}
\end{center}
\caption{Scaling laws in joint functional/evolutionary partitioning of
  genomes.  Genomes are partitioned into families of homologous genes
  (colors) and functional categories (shapes).  (A) The number of
  unique evolutionary categories (domain families) (y-axis) scales
  sub-linearly with the genome size (x-axis.) (B) Cumulative
  histograms of domain family size (see Figure~\ref{fig:histograms}).
  (C) The number of transcriptional regulators (red), metabolic
  enzymes (blue), and housekeeping genes responsible for translation
  (green) plotted as a function of the genome size measured by the
  total number of domains.  Symbols in all the plots are empirical
  data for protein domains in $ 753 $ fully sequenced bacterial
  genomes.  }
\label{fig:partitioning}
\end{figure}

Protein-coding genes in genomes can be classified in both functional
categories (e.g. transcription factors or metabolic enzymes) as well
as ``evolutionary categories'' or families of homologous
genes 
(to avoid confusion, in the following we will reserve the term
``category'' to functional annotations, and we will use the term
``family'' as a generic indication of homology classes, or domain
families/superfamilies in domain data, see Methods).
Functional categories are routinely composed of a large number of
evolutionary ones. This distinction is illustrated in
Fig.~\ref{fig:partitioning}, where genes are characterized by both
shape (functional category) and color (homology class) with each shape
represented by multiple colors.
Understanding the principles connecting these separate classifications
of genomic material is an important step in order to disentangle the
organization of the content of whole genomes.

More specifically, studies of fully sequenced genomes revealed that
their functional and evolutionary composition is governed by simple
quantitative laws~\cite{HvN98,vannim_03}.  In particular, for
prokaryotes the number of genes in individual functional categories
was shown to scale as a power law of the total number of genes in the
genome \cite{vannim_03}.  Depending on the functional category the
exponent of this scaling law varies from 0 (for fixed sets of
housekeeping genes) to 1 (for metabolic enzymes) and all the way up to
2 (for transcription factors and kinases)~\cite{vannim_08,vannim_03}.
Furthermore, the distribution of sizes of gene families (called
``evolutionary categories'' in our title) has a scale-free
distribution with the exponent inversely correlated with the genome
size~\cite{HvN98,KWK02,DSS02}.  The overall number of gene (or domain)
families represented by at least one member exhibits a
slower-than-linear scaling with the total number of genes in a
genome~\cite{bassetti,messico}.
\rev{
  Biologically, the growth of evolutionary families derives from
  combined processes of horizontal gene transfer, gene duplication,
  gene genesis, and gene loss~\cite{Koonin}. For prokaryotes,
  horizontal transfer appears to dominate gene family
  expansion~\cite{Treangen2011a}, and the same process is presumably
  very important for the introduction of a new evolutionary family
  into an extant genome.
}

The comprehension of these empirical laws requires to construct
quantitative models that explore different design principles, or more
prosaically the recipes by which genomes are built from elementary
functional and evolutionary ingredients.  In this study we introduce
\rev{the first model to jointly explain observed scaling laws for
  evolutionary families and functional categories}.

Several theoretical models have been previously proposed to explain
the observed power-law distribution of family
sizes~\cite{KWR+02,QLG01,KLQ+06,DS05,DSS02} Most of these models are
of \rev{class-expansion}/innovation/loss type\rev{, abstractly
  mimicking basic evolutionary moves such as horizontal transfer,
  duplication, loss}. We recently formulated a related model that in
addition to family size distribution also explains and successfully
fits the scaling of the number of distinct gene families represented
in a genome as a function of genome size~\cite{bassetti,angelini}.

On another front, the ``toolbox model'' of evolution of metabolic
networks and their regulation recently proposed by one of
us~\cite{maslov} offered an explanation for the quadratic scaling
between the number of transcription factors and the total number of
genes in prokaryotes.
In this model, metabolic and regulatory networks of prokaryotes are
shaped by addition of co-regulated metabolic pathways. The number of
added enzymes systematically decreases with the proportion to which
the organism has already explored the universe of available metabolic
reactions, and thus, indirectly, with the size of its genome.
For the purposes of the present study, a key ingredient of the toolbox model
is that events adding or deleting genes in multiple functional
categories (in this case metabolic enzymes and transcription factors
regulating metabolic pathways) are tightly correlated with each other.
The concept of coordinated expansion or contraction of functional
categories can in principle be extended beyond enzymes and their
regulators.

One should note that this explanation of scaling of functional
categories is conceptually different from that based on ``evolutionary
potentials'' proposed in Ref.~\cite{vannim_08}.  Evolutionary
potentials quantify the intrinsic growth rates of
individual categories.
\rev{ This means that in this model the growth of one functional
  category is represented as uncoupled from growth or decline in other
  functional categories.
  However, evolutionary potentials could also be the effective result
  of the coordinated expansion of multiple functional categories
  linked by interactions of biological and evolutionary origin
  (e.g. linking membrane proteins with signal transduction, etc.)
  On the other hand, it is clear that models with evolutionary
  potentials represent quite well the empirical data on the growth of
  functional categories, and thus it appears that this must be (at
  least) a very good effective description, that any more detailed
  model needs to reproduce. 
}

This study brings together the basic ingredients of
\rev{class-expansion}/innovation/loss models~\cite{bassetti,angelini}
and coordinated growth of functional categories~\cite{maslov}. The
resulting combination allows us to study the interplay between the
scaling of evolutionary and functional categories.  In particular, we
mathematically derive a relation between the exponents characterizing
these two scaling laws.  It predicts that functional categories that
grow faster-than-linearly with genome size are characterized by
flatter-than-average family size distributions. This prediction of our
model is subsequently verified by our analysis of functional and
evolutionary scaling in empirical data on sequenced prokaryotic
genomes. \rev{Finally, we analyze and discuss the alternative
  combination of a class-expansion/innovation/loss model with growth
  of functional categories dictated by evolutionary potentials. }

\section{MATERIALS AND METHODS}

\subsection*{Models}

The model represents a genome as a list of genes, which is partitioned
in homology families and functional categories.  Genome evolution is
modeled as a stochastic process where the elementary moves can be any
of two types: (i) a ``family expansion'' or ``duplication'' move in
which a new domain is placed in an evolutionary category (family of
homologous domains) already present in the genome or (ii) an
``innovation'' move in which a new family with just one domain appears
in a genome (e.g. by the virtue of horizontal gene transfer).

We would like to emphasize that \rev{in the tradition established in
  ``duplication-innovation-loss'' models, which we follow,} the family
expansion move \rev{is customarily labeled} as duplication. In reality
this move can come either by the virtue of gene duplication or by
horizontal gene transfer, which appears to be the dominant
class-expansion mechanism in bacteria~\cite{Treangen2011a}.  The
overall family size in all genomes might be generating an effective
``preferential attachment'' for HGT events (see
Refs.~\cite{isambert,vannim_08} and open comments by referees
therein).

Although genes are natural objects of this kind of
description, it is not simple to use genes as central units in the
analysis of empirical data, mainly due to the fact that gene dynamics
is complex and may contain events of gene fusion, splitting and
internal rearrangements. Thus, as in some previous analyses, we will
compare the models with data on protein
domains~\cite{vannim_08,bassetti}, which have the important property
that they cannot be split into smaller units~\cite{BBK+05}.  Domains
are modular building blocks of proteins and it has been argued that
they effectively work as the natural atomic elements in genome
evolution~\cite{KWK02}. Concerning the scaling laws, domains appear to
have the same qualitative behaviour as genes. Throughout the paper, we
will be comparing the models with data on 753 bacteria from the
SUPERFAMILY database~\cite{supfam_funz}. The models will be formulated
for abstract atomic elements that could be genes or domains, and
possible relevant issues when dealing empirically with genes will be
addressed in the discussion. In describing the models we will
generally refer to these units as genes.

\begin{figure}[t]
\begin{center}
  \includegraphics[width=0.9\textwidth]{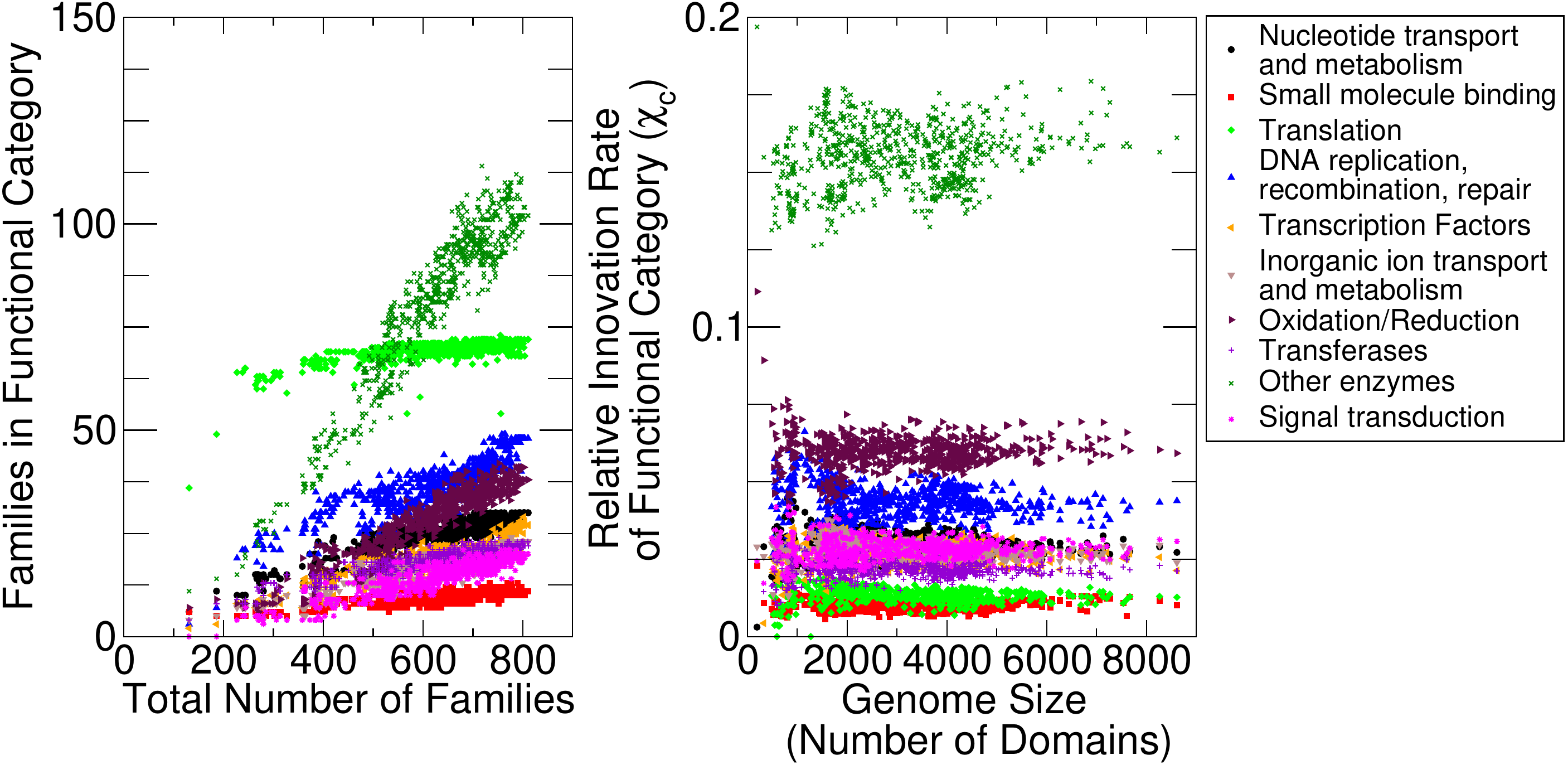}
\end{center}
\caption{The number of evolutionary (domain) families belonging
    to a functional category follows a linear law in empirical data,
    denoting a possible invariant of genome composition. The left
    panel plots the data on for the number of families $f_c$ in the
    ten largest functional categories on all genomes, following the
    trend $f_c = A_c + \chi_c f$, where $f$ is the total number of
    families on the genome.  Symbols are empirical data for $ 753 $
    fully sequenced bacterial genomes. The offset $A_c$ is large only
    for the ``translation'' category.  The right panel is a plot of
    the coefficients $\chi_c$ obtained from the same data (subtracting
    the offset $A_c$ obtained from a linear fit), as a function of
    genome size in domains, $n$. See also Supplementary
    Table~S1. }
\label{fig:Chi_c}
\end{figure}

Technically, in order to compare with the protein domain data we rely
on simplifying assumptions on the domain composition of proteins.
Obviously the situation is more complex than this.  We have verified
in the data that the number of TF domains are linear in the number of
TF genes (Supplementary Figure~\ref{fig:supp-gene}), with slope $
1.09 $ (average number of TF domains in a TF gene).
A second assumption is that the number of families belonging to a
functional category is linear in the total number of families. This
assumption is in accordance with data (see Figure~\ref{fig:Chi_c} and
Supplementary Figure~\ref{fig:supp-vannimSF}).  In particular, we
observed this trend for the number of transcription factor
superfamilies (see Supplementary Figure~\ref{fig:supp-bpTFsupfam}).

\subsubsection*{Standard Chinese Restaurant Process.}

The starting point is a \rev{class-expansion}/innovation process for the
homology families that reproduces qualitatively the empirical scaling
laws~\cite{bassetti}. This process (known in mathematical literature
as ``Chinese Restaurant Process''or CRP~\cite{pitman}) defines
a growth dynamics for the partitioning of a set of elements (genes or
domains) based on two basic growth moves.
Traditionally the CRP model is defined by two parameters $ \alpha $
and $ \theta $ constrained by $ 0 \leq \alpha \leq 1$ and $ \theta > - \alpha $.
The moves are quantified and
defined by two probabilities $ p_O $ and $p_N $ of duplication and
innovation respectively.
\begin{itemize}

\item The \rev{class-expansion} probability $p_O^i$ of a domain family
  $i$ is proportional to the number of family members $n_i$ currently
  in the genome offset by $\alpha$: $p_O^i \sim n_i-\alpha$ (see
  Table~\ref{tab:notation}).

\item The innovation probability $ p_N $ is the probability of adding
  a new domain family with one member. It corresponds to a new domain
  family appearing in a genome by \textit{de novo} evolution or
  horizontal gene transfer.  The CRP model assumes $p_N \sim \alpha f
  +\theta$, where $ f $ is the total number of domain families present
  in the genome.
\end{itemize}
The normalization condition $ p_N+ \sum_i p_O^i=1 $ determines the
pre-factor in both equations to be $1/(n+\theta)$.  A gene loss move
does not seem to be essential for the basic qualitative
results. Indeed, if stochastic (uniform) gene loss is
incorporated into the model it results only in renormalization of
parameters $ p_O $ and $p_N $~\cite{angelini}.

We explore the model by direct simulation and by solving continuous
``mean-field'' equations~\cite{bassetti,angelini} that describe the
mean behaviour of the number of homology families and functional
categories, and the statistics of the population of families and
categories.

\subsubsection*{CRP model incorporating functional categories.}
In order to introduce functional categories into the CRP, one has to
specify $ p_O $ and $ p_N $ for different categories.
We first assume that the probability of introducing a gene of a
specific functional category by the innovation move is independent of
genome size. This assumption implies that the number of homology
families of a given category scales linearly in the total number of
families, and is justified empirically for some functional categories
by domain data (see \rev{Figure~\ref{fig:Chi_c} } and Supplementary
Figures~\ref{fig:supp-vannimSF} and~\ref{fig:supp-bpTFsupfam}).
Equivalently, $p_N^c = \chi_c p_N$, where $
\chi_c $ is the probability of introducing a new family of the
category $ c $.
\rev{In other words, it is assumed here that every time a new family
  is added, the probability that it will belong to category $c$ is
  $\chi_c$.}

Under this assumption, the mean-field equation describing the growth
of a family of homologous domains (evolutionary category) is
\rev{
\begin{equation}
  \displaystyle
  C(n) \partial_{n} n_i = \displaystyle 
                          \sum_{j=1}^{f}  a_{ij} n_j  \,  - \alpha.
\label{eq:CRPdupl-class}
\end{equation}
} 
Here the genome size $n$ is used instead of time and averages over
multiple realizations of a process are implied.  The novel ingredient
of the model - coordinated growth of functional categories - is
encoded in the coefficients $a_{ij}$ responsible for correlated
duplications between evolutionary families $i$ and $j$.  We assume
$a_{ij}$ to depend only on functional roles of families $i$ and $j$.
The equation describing the growth of $f$ - the number of distinct
families in a genome is the same as in a standard CRP model.
\begin{equation}
C(n) \partial_{n} f = \displaystyle (\alpha f + \theta) \ .
\label{eq:CRPduplcorrFtot}
\end{equation}
The function $ C(n) $, which sets a natural time scale for the
process, is determined by the normalization condition $\partial_n n =
1$, i.e.  $\sum_i \partial_n n_i + \partial_{n} f = 1 $.

For the specific case of categories of transcription factors (TFs)
regulating metabolic processes and their metabolic target enzymes, the
necessity of a correlated move can be argued along the lines of
Ref.~\cite{maslov}. A set of new targets has to be added to
incorporate a new metabolic function.
%
This entails the addition of a new metabolic pathway that is long
enough to connect a new nutrient to a previously existing pathway,
that further converts it to a central metabolic ``core network''.
Supposing that each newly added branch is controlled by only one added
transcription factor, since the length of the branch becomes smaller
with increasing size of the organismal metabolic network (compared to
a metabolic ``universe''), on average, increasingly more TFs per
target will be needed in order to control newly incorporated branches.

More generally, functional, genetic and epistatic interactions can
create the correlated growth of different functional categories of
genes. In the discussion section we provide the empirical evidence of
statistically significant correlations between various functional
categories.

Following the recipe outlined in Ref.~\cite{maslov} \rev{we consider}
a simplified version of the model involving only two functional
categories: 1) $TF$- transcription factors controlling metabolic
processes; 2) $met$ - metabolic enzymes they regulated. 
\rev{ As in the toolbox model, changes in $n_{TF}$ and $n_{met}$ are
  coordinated with correlation coefficients $a_{ij}$ given by
\begin{eqnarray*}
  \displaystyle
  a_{ij} = \frac{n_{met}}{U}\, ,  \  a_{ji}=0 ;   \
  \mathrm{for} \ i\ne j \\
  \mathrm{and} \ a_{ii} = 0 \  .  
\end{eqnarray*}
Here $U$ is the size of the metabolic universe, $i$ denotes any gene
family from functional category $TF$, and $j$ - from the functional
category $met$. In this variant, addition of transcription factors can
only occur conditionally to the addition of metabolic enzymes. In the
following, we will refer to this model variant as model Ia.
} 
\rev{ We define a second variant of the correlated model (model Ib),
  which is a more direct extension of the standard CRP model, and thus
  can exploit previous mean-field theory analytical results. In this
  case
\begin{eqnarray*}
  \displaystyle
  a_{ij} = \frac{n_i}{n_{met}}\, , \   a_{ji}=0 ;   \
  \mathrm{for} \  i\ne j \\
  \mathrm{and} \  a_{ii} = 1 \  ,  
\end{eqnarray*}
(where $i$ again denotes any gene family from functional category $TF$
and $j$ - from the functional category $met$).  In this model variant,
all families (and hence also transcription factors families) have an
equal intrinsic growth rate on top of the correlation. If $a_{ij} = 0,
\ i\ne j$ the model is equivalent to the standard CRP.
}
%
\rev{Finally, we also considered a model (model II)} where
correlations between functional categories are absent, but instead
members of a given functional category are added at a
category-dependent intrinsic rate as prescribed by ``evolutionary
potentials'' of Molina and van Nimwegen (in this case, $a_{ij} = 0$
for $i \ne j$, and $a_{ii} = \rho_{c(i)}$, where $c(i)$ is the
functional category to which family $i$ belongs, and $\rho_{c(i)}$ is
the evolutionary potential of class $c$).
These results are discussed later on in the manuscript and compared to
to the two ``correlated duplication'' models above (see Discussion and
Supplementary Text).

\rev{ 
  To resume, two kinds of models are considered here: ``correlated
  recipes'', where the scaling exponents can only result from
  interactions between categories (model Ia and Ib, the main focus of
  our study), and ``absolute recipes'' (model II), leading to
  different intrinsic growth rates for different
  categories. Correlated models might contain an specific intrinsic
  growth rate of the classes, equal for all classes (model Ib), or not
  (model Ia).  We will see that the important distinction between
  model I (a and b) and model II is that the different scaling
  exponents for functional categories are a result of correlations and
  not absolute class expansion rates.  }

\subsection*{Data}
Data on superfamily domain assignments and superfamily functional
annotations for the $ 753 $ Bacteria were obtained from the
SUPERFAMILY (v1.73) database~\cite{supfam_funz}.  The database
contains $1291$ different domain superfamilies grouped into $ 47 $
different functional categories ($ 60 $ families do not belong to a
specific category). These categories are divided into $ 6 $ larger
groups (Metabolism, General, Regulation, Information, Initiation
Complex Processes and Elongation Complex Processes, see also
\\http://supfam.cs.bris.ac.uk/SUPERFAMILY\_1.73/function).

\rev{
\subsection*{Evaluation of exponents in empirical data}
We considered the normalized cumulative histograms (families with more
than $d$ members) and non-cumulative histograms (families with exactly
$d$ members) of the populations for all evolutionary families (related
to exponent $\beta$, see Results), and those restricted to the families
belonging to each of the main functional categories indexed by $c$
(related to the exponent $\beta_c$, see Results).  Exponents were
estimated by fitting the data with a power-law, restricting to a
window where the $x$ axis value was less than a cutoff value, as in
Ref.~\cite{angelini}. The cutoff was chosen for each fit, by
minimizing the chi square residuals with varying window size.  This
procedure was implemented with a custom CINT (C++) script using the
ROOT software.  Figure~\ref{fig:expcorr} is obtained considering the
fitted exponents for the histograms of the five largest genomes (where
the ``finite-size correction'' is smallest, see Figure~\ref{fig:histoexp}
and Ref.\cite{angelini}.)
}

\subsection*{Empirical correlations among functional categories}
Correlation between families (or categories) populations were
calculated from the deviations from the average trend.  We obtained
the frequency of a family/category in every genome, defined as the
ratio between the population of a family in domains and the total
number of domains assigned on that genome.  Subsequently for every
family/category, we extracted an average trend as a function of genome
size $n$ using a sliding-window histogram (with window size of 280
domains and resolution of 28 domains), and we considered the deviation
of each genome from the average trend at its value of $n$.  The
Pearson correlation of these deviations over all the genomes was
considered between each pair of families/categories \rev{(Figure
  \ref{fig:correlation} and Supplementary
  Table~S3 and~S4)}.

\subsection*{Models and simulations}
The quantitative duplication-innovation evolutionary models were
explored by a mean-field analytical approach and direct numerical
simulations. The mean-field approach considers equations for the means
of the observed quantities in the large-$n$ approximation.
In parallel with the mean-field analysis, we performed simulations of
the main model and its variants.  The realizations depend on the
following parameters. (i) The parameters of a standard CRP, $ \alpha $
and $ \theta $. (ii) The parameter $ \chi_c $, i.e. the probability
that a new family belongs to a given functional category.  This
parameter can be inferred from data \rev{(see Results and
  Figure~\ref{fig:Chi_c})}.  For example, for the case of
transcription factors and targets, we defined $\chi_{TF} $ from the
slope extrapolated from Supplementary
Figure~\ref{fig:supp-bpTFsupfam}, giving $ \chi_{TF} \simeq 0.035 $
(see also Supplementary Figure~\ref{fig:supp-simTFsupfam}). (iii)
Initial conditions, represented by initial configuration (number of
leaves, number of TFs and number of families in both categories). We
have used the configuration of the smallest bacterium in the dataset
(Candidatus Carsonella ruddii).  An alternative choice could be the
minimal intersection of all genomes in the database. (iv)
Variant-specific parameters, that amount to the evolutionary
potentials $\rho_c$ for the first variant of the model, and the
correlation matrix between functional categories, $a_{ij}$ for the
second variant.  Simulation results are typically visualized in
boxplots in order to compare the means with the probability
distributions. In these plots bars correspond to (in order) the
smallest observation, lower quartile, median, upper quartile, and
largest observation.

\section{RESULTS}

\rev{
\subsection*{A new invariant of genome composition}
We found (Figure~\ref{fig:Chi_c} and Supplementary Table See also
Supplementary Table~S1)
that the number of evolutionary domain families forming a functional
category follows a linear law in empirical data, denoting a possible
invariant of genome composition.  This also implies that the mean law
$\partial_n{f_c} = \chi_c \partial_n{f}$ assumed in the model is
justified by the data.  This does not mean exactly that the fraction
of all families belonging to a certain functional category is
constant. Rather, the observed law can be $ f_c = A_c + \chi_c f$,
with an offset $A_c$ representing a minimal amount of evolutionary
families required to build a given functional category. In empirical
data, this offset appears to be large only for the ``translation''
functional category.
}

\subsection*{The model captures the combined scaling laws}

\begin{figure}[t]
\begin{center}
\includegraphics[width=0.9\columnwidth]{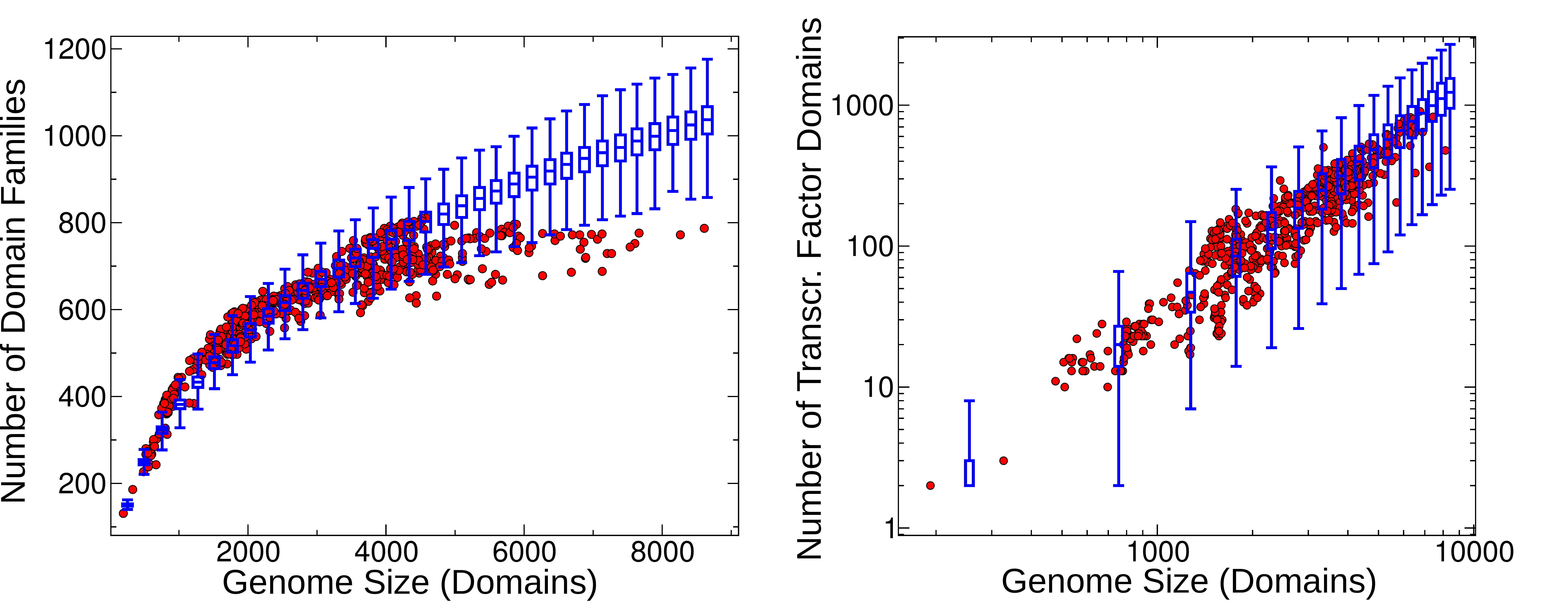}
\end{center}
\caption{Comparison between $ 1000 $ realizations of the correlated
  duplication model \rev{Ib} at $ \alpha = 0.3 $ and $ \theta = 140 $
  (blue boxplot) with empirical data (red). The left panel is a plot
  of the number of distinct domain families versus genome size.  The
  fact that the number of families does not saturate is a property of
  the standard duplication-innovation model (see~\cite{bassetti} for a
  complete discussion). The right panel plots the number of TF domains
  versus the total number of domains, showing that the scaling of the
  transcription functional category is well reproduced (exponent $
  \simeq 1.6 $.) \rev{See Supplementary Figure~\ref{fig:supp-ToolPan}
    for model Ia.} }
\label{fig:corresults}
\end{figure}

Numerical simulation and mean-field analytical solutions of the
correlated growth model \rev{ (model I) reproduce very well both the
  empirical behavior of the TFs scaling law and the statistics for
  evolutionary domain families (Figure~\ref{fig:corresults} and
  Supplementaty Figure \ref{fig:supp-ToolPan}). We found no
  significant qualitative difference between models Ia and Ib
  regarding these observables. Furthermore, the joint scaling laws can
  be reproduced also with an uncorrelated model (model II), with minor
  technical difficulties (see Discussion). }
\rev{ The correct asymptotic quadratic scaling can be obtained from
  mean-field arguments for both model I and II. These arguments are
  presented in the Supplementary Text. In order to illustrate this
  point we consider for example model Ib.  Starting from
  Eqs.~\ref{eq:CRPdupl-class} one has to sum over all domain families
  from functional categories $TF$ and $met$.  Since $n_{TF} = \sum_{i
    \in TF} n_i$, depends on the number of TF classes, one must have
  for its derivative $ \partial_n n_{TF} = \sum_{i \in TF} \partial_n
  n_i + \partial_n f_{TF}$. Combined, these two equations give $d
  n_{TF}/d n_{met} = 2(n_{TF}-\alpha)/(n_{met}-\alpha) \simeq
  2n_{TF}/n_{met}$, or finally the quadratic scaling $n_{TF} \sim
  n_{met}^2$.  }

\rev{Altogether, the agreement between data and model} is universal,
in the sense that the same three parameters are sufficient to predict
family/category numbers and populations for all genomes in the
dataset. Moreover, the comparison does not rely on the adjustment of
any hidden parameter.
It is also worthwhile noting that, while the input of \rev{model I (a
  and b)} is built to give an \rev{asymptotic} power-law scaling
exponent of two for transcription factors (which is reproduced by the
mean-field approach), at the relevant genome sizes the model
automatically reproduces the \emph{correct} empirical exponent (about
$1.6$ in the SUPERFAMILY data) \rev{as an effect of the finite system
  size}.
Note that \rev{in model Ib TFs can duplicate both spontaneously
  (uncorrelated move) and following spontaneous duplication of targets
  (correlated move), corresponding to the terms $a_{ii}$ and $a_{ij}$
  in equation~\ref{eq:CRPdupl-class}, while in model Ia this does not
  happen. }

The extension of the model to more than two categories requires to
know the laws through which families of different categories are
correlated with each other. Supplementary
Figure~\ref{fig:supp-comparision} compares the results obtained by a
correlated duplication model formulated with three categories (TFs,
met, others).
%

\subsection*{Prediction of the exponents of the family-population
  histogram restricted to single functional categories.}

\begin{table}[b]
\caption{Basic model quantities and notations}
\label{tab:notation}%
\resizebox{\columnwidth}{!}{
\begin{tabular}{|c|c|}
\hline
Quantity & Meaning \\
\hline
$ \alpha $ , $ \theta $ & CRP model parameters \\
\hline
$ n $ & Genome size quantified by its total number of domains \\
\hline
$ n_i $ & Number of domains in the family $ i $ \\
\hline
$ n_c $ & Number of domains in the functional category $ c $ \\
\hline
$ f(n) $ & Number of families in a genome of size $ n $ \\
\hline
$ f_c(n) $ & Number of families in a genome of size $ n $ belonging to the functional category $ c $ \\
\hline
$ f(d,n) $ & Number of families with exactly $ d $ members in a genome of size $ n $ \\
\hline
$ f_c(d,n) $ & Number of families belonging to the functional category $ c $ with exactly $ d $ members in a genome of size $ n $\\
\hline
$ \beta $ & Exponent of the family-population histogram \\
\hline
$ \beta_c $ & Exponent of the family-population histogram restricted to category $ c $ \\
\hline
$ \chi_c $ & Probability to introduce a new family of the category $ c $
	(empirically quantified by the slope of $ f_c $ vs. $ f $) \\
\hline
$ \zeta_c $ & Exponent of the scaling of the size of functional category $ c $ vs. genome size $ n $\\
\hline
\end{tabular}%
}
\end{table}

\begin{figure}[tp]
\begin{center}
\includegraphics[width=0.75\columnwidth]{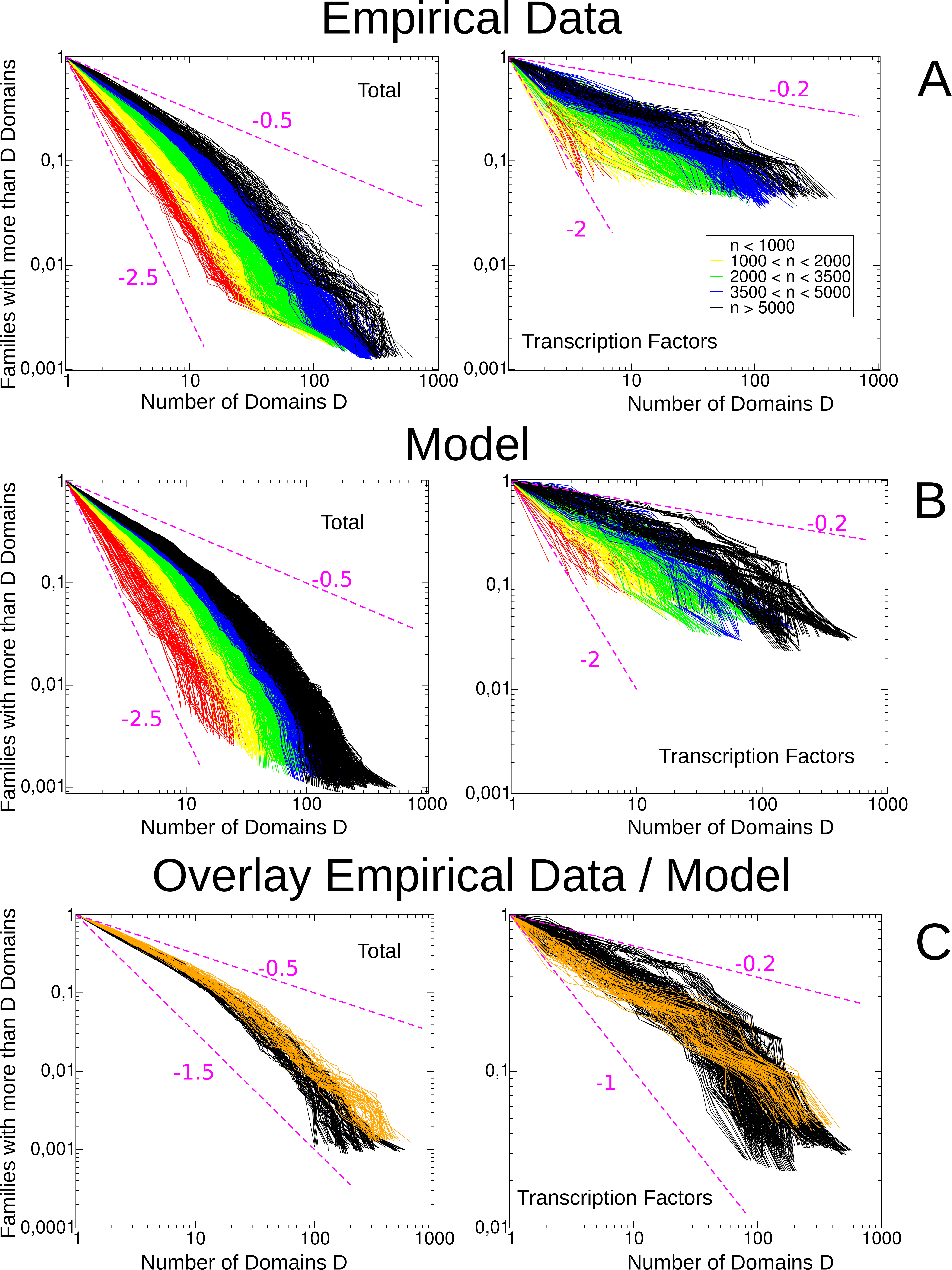}
\end{center}
\caption{ Empirical data and simulations for the normalized domain
  family population cumulative histograms. The histograms are defined
  as the fraction $f(d,n)/f(n)$ of families with more than $d$
  domains.  (A) Empirical data for the 753 bacteria in the SUPERFAMILY
  database (each color is a set of genomes with similar sizes).  Left
  panel: domain family population cumulative normalized
  histograms. Right panel: normalized cumulative histograms restricted
  to domain families belonging to the transcription factor functional
  categories. Note that the histograms slopes are different.  (B)
  Simulations for domains family population cumulative histograms of
  CRP with correlated duplications run at $ \alpha = 0.3 $ and $
  \theta = 140 $.  \rev{The plots in the two panels are defined as in
    (A)}.  (C) Comparison between simulations of the correlated
  duplication model variant run at $\alpha = 0.3 $ and $\theta = 140 $
  (black lines) with empirical data (orange lines) for the largest
  genome sizes ($ 5000 < n < 8500 $).  Left panel: global normalized
  cumulative histograms of domain family population.  Right panel:
  normalized cumulative histograms restricted to transcription factor
  domain families.  }
\label{fig:histograms}
\end{figure}

While the agreement between model and data shows that the scaling of
functional and evolutionary categories can be understood jointly, it
does not provide by itself any substantially new information about how
the two partitionings interact. Further insight can be obtained
considering the 
\rev{distributions of the number of domains per family for different
  evolutionary families belonging to the same functional category}.
In general, the population of domain families of a genome follows a near power
law distribution whose slope depends on genome size
(Figure~\ref{fig:histograms}). The mean number $f(d,n)$ of domain
families having $d$ members at large genome size $n$ is well described
by the slope $ 1/d^{1+\beta} $ (see Figure~\ref{fig:histograms}) ,
and thus the cumulative histogram by $Q(d,n) \sim 1/d^{\beta}$
, where the fitted exponent $ \beta $ typically lies between 0 and 1.
The standard CRP predicts this behavior~\cite{bassetti,angelini}.
The model described here here allows to consider the same histograms
restricted to specific functional categories (Figure
\ref{fig:histograms} and Figures \ref{fig:expcorr}).

\begin{figure}[t]
\begin{center}
\includegraphics[width=0.9\columnwidth]{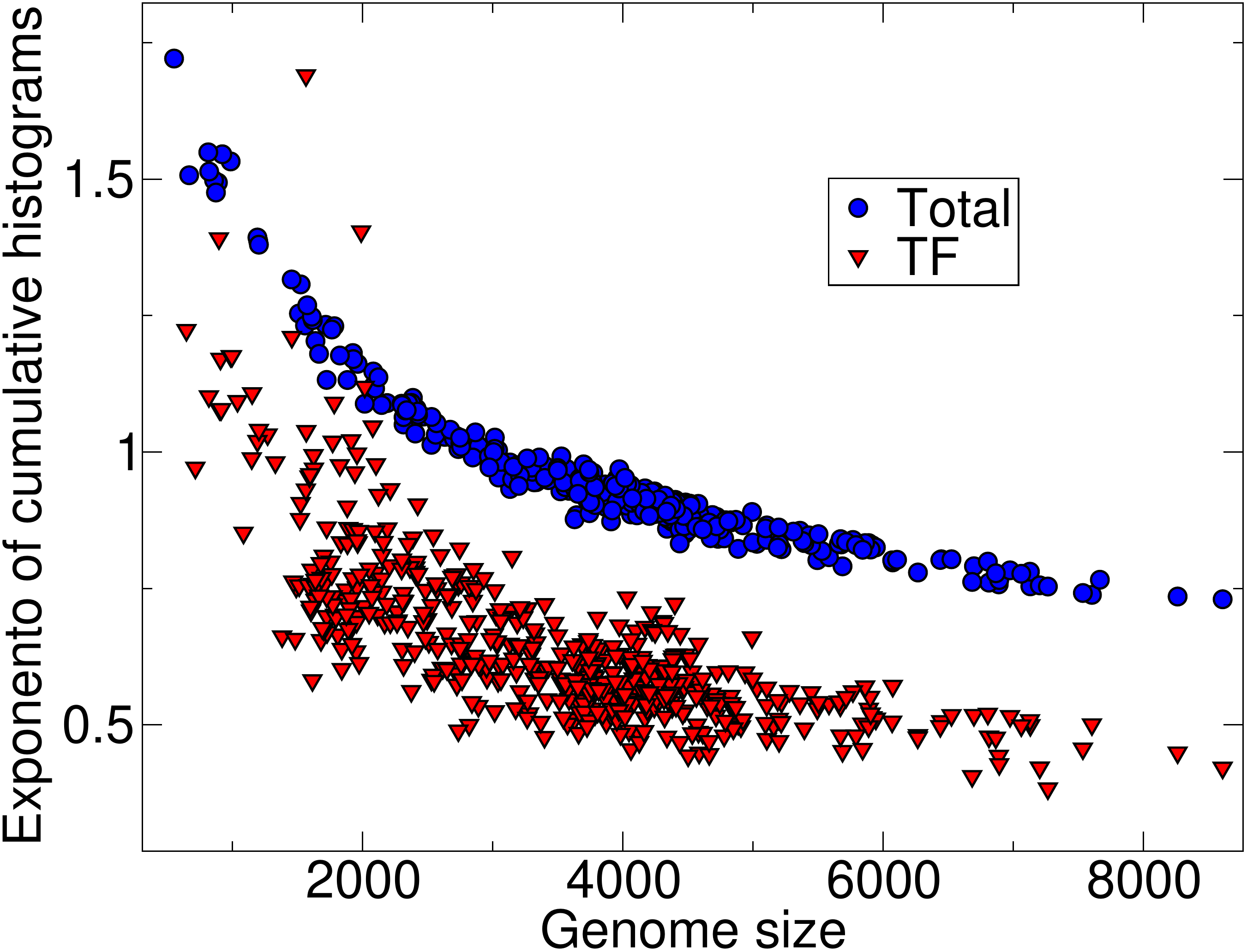}
\end{center}
\caption{ Exponent of evolutionary families and genome size.  Fitted
  exponent of domain family population cumulative histograms vs.
  genome size, for the 753 bacteria in the SUPERFAMILY database for TF
  families (red circles) and all families (black squares), obtained by
  a fitting method giving a lower weight to the tail in order to keep
  into account the cutoffs (used in Ref.~\cite{angelini}).}
 \label{fig:histoexp}
\end{figure}

A mean-field calculation (see Supplementary Text) based on the model
variant with correlated duplication predicts that the different trend
of domain population histograms for transcription-factor families
scales as $f(d,n)_{TF} \sim 1/d^{1+\frac{\beta}{2}} $ (see
Figure~\ref{fig:histoexp}).
Thus, the ratio between the exponent of the cumulative histogram of
all families and the exponent of the cumulative histogram restricted
to families belonging to the transcription factor category is
predicted to be equal to the mean-field exponent for the scaling of
the functional category.  
\rev{ Specifically, $Q(d,n)$ scales as $1/d^{\beta}$ whereas
  $Q_{TF}(d,n)$ scales as $d^{-\beta/2}$ and thus the ratio of
  exponents is $\beta/(\beta/2) = 2$, and this matches the asymptotic
  scaling of the number of transcription factors.  }
%
%
\rev{ More in general, the model indicates that each time the
  per-family duplication probability for a functional category takes
  the form $ p_{O}^c \simeq \zeta_c n_{c}$, where $n_c$ is the total
  population of the functional category $c$, the coefficient $\zeta_c$
  will appear in the equation for $P(d)_{c}$, the (cumulative)
  distribution of families belonging category $c$.  This causes the
  relationship $\beta_c = \beta / \zeta_c$ and appears to be robust
  with respect to the choice of a specific model (see Supplementary
  Text).}
In other words, a precise quantitative relationship must exist between
the scaling exponent of a category and the slope of the family
population histogram restricted to the same category.  Functional
categories that grow faster-than-linearly with genome size will have
flatter-than-average domain family size distributions. Conversely
categories growing slower-than-linearly will follow a
steeper-than-average slope.

Accordingly, a strongly visible trend should be expected in empirical
data from families belonging to the transcription factor category,
which scales with exponent 2.  Indeed, the empirical population
histograms for the transcription factor functional category for all
the genomes in the data set have a slope that is spectacularly
different from the global one (Figure~\ref{fig:histoexp} \rev{and
  Supplementary Figure~\ref{fig:supp-scatterExp}}).
Quantitatively, this observation is in excellent agreement with
predictions (Table~\ref{tab:exphistotf}).
Direct simulations of the correlated model reproduce well both the
behavior of the histograms at given size and the dependency on genome
size (see Figure~\ref{fig:histograms}).

More generally, one can test the prediction $\zeta_c = \beta /
\beta_c$ with an empirical evaluation of many functional categories
(Figure~\ref{fig:expcorr}). The agreement of empirical data with the
predicted behavior is reasonably good, keeping in mind that many
functional categories are composed by few or poorly populated families,
and in these cases the data might not follow a scaling law that is as
clearly defined as the metabolism or the transcription factor
categories.

\begin{figure}[!ht]
\begin{center}
\includegraphics[width=0.9\columnwidth]{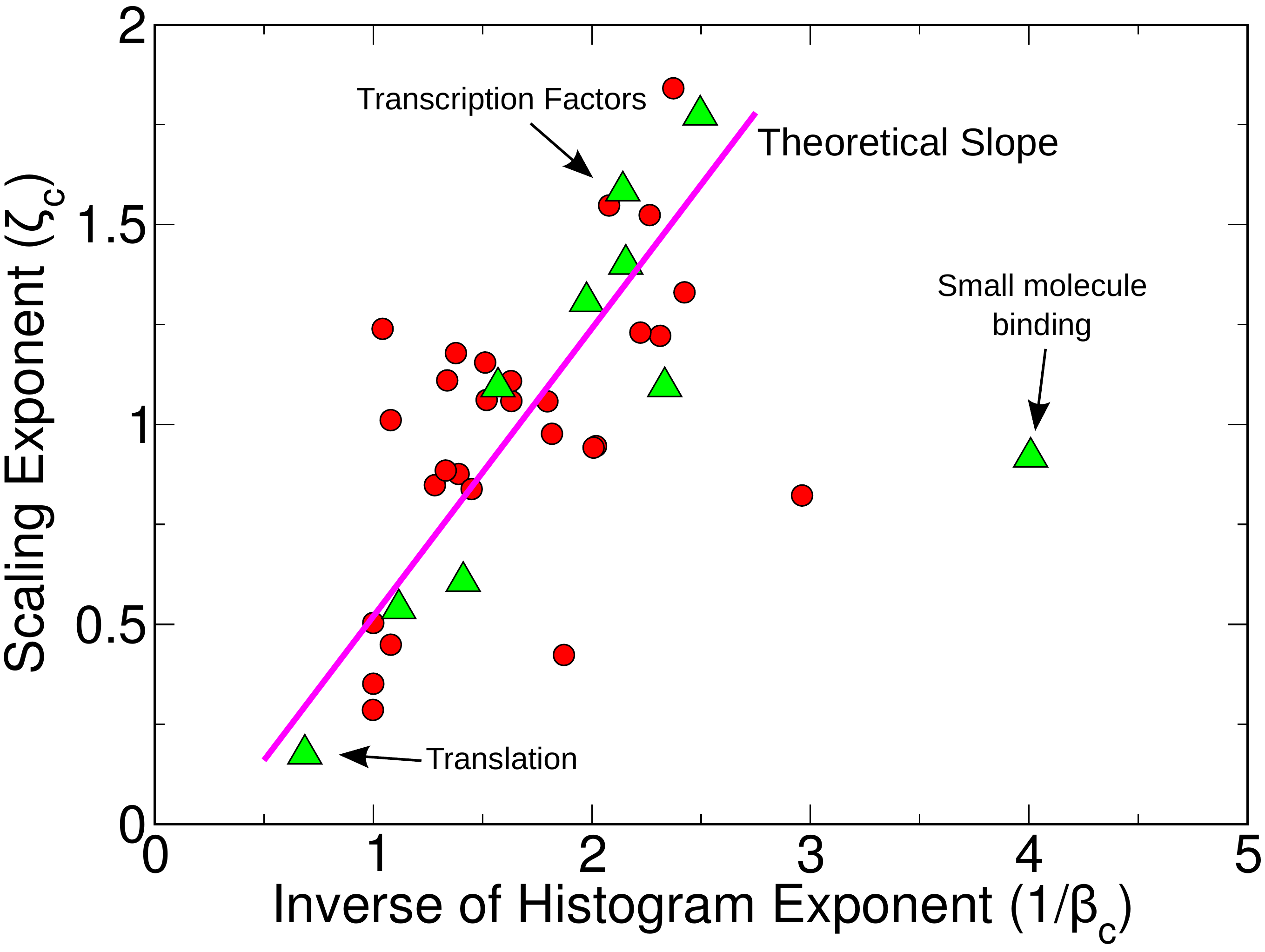}
\end{center}
\caption{ Linear relation between $\zeta_c$ and $1/\beta_c$.  Our
  theory predicts $\zeta_c \sim \beta/ \beta_c$ (solid line). The
  empirical value of $ \beta = .74$ is calculated from the family
  population histograms of the five most populated genomes.  Symbols
  (circles and triangles) are empirical data for $ 38 $ functional
  categories \rev{(see also Supplementary Table~S2)}. Triangles represent the ten most populated
  categories, where the estimated exponents are most accurate.  The
  outlier is the ``small molecule binding'' category known to follow
  peculiar evolutionary mechanisms~\cite{Koonin01}.  }
  \label{fig:expcorr}
\end{figure}

\begin{table}[b]
  \caption{  Prediction of the exponent of the
    family-population histograms restricted to singular functional category.
    Comparison between expected and
    observed ratio of the exponent of the cumulative histogram of all
    families and the exponent of the cumulative histogram of
    transcription-factor families (
    Figure~\ref{fig:expcorr} ), for the five largest bacteria in
    the SUPERFAMILY database. The ratio can be compared with the
    mean-field prediction of $ 2 $, or directly with the empirical exponent
    of the transcription factor functional category ($ 1.6 $). }
 \label{tab:exphistotf}
 \begin {tabular}{|c|c|c|}
\hline
   Genome  & $\beta/\beta_{TF}$  & $\zeta_{TF}$ \\
   \hline 
   Sorangium cellulosum     & $ 1.72 \pm 0.1 $  & $ 1.6 $ \\
   \hline
   Burkholderia xenovorans  & $ 1.63 \pm 0.08 $ &  $ 1.6 $ \\
   \hline
   Burkholderia             & $ 1.54 \pm 0.13 $ & $ 1.6 $ \\
   \hline
   Solibacter usitatus      & $ 1.46 \pm 0.05 $ &  $ 1.6 $ \\
   \hline
   Bradyrhizobium japonicum & $ 1.59 \pm 0.11 $ &  $ 1.6 $ \\
   \hline
 \end{tabular}
 \end{table}


\section{DISCUSSION}

 \subsection*{Population of evolutionary families of a given functional
   category}

We have presented the first combined quantitative description of the
partitioning of genomes in both evolutionary families and functional
categories.  The results show that a theoretical framework that
correctly reproduces both the scaling laws for functional categories
of genes/domains and the scaling laws for gene/domain families
(numbers and histograms) is possible. 
Biologically, this finding can help us understand the large-scale
architecture of a genome in terms of its functional content.

\rev{ Analyzing the data in order to formulate the model, we found
  that the number of evolutionary domain families forming a functional
  category is linear in the total number of domain families
  (Figure~\ref{fig:Chi_c}). Thus, the genomic subdivision of
  evolutionary classes in functional categories appears to be arguably
  the simplest possible, if one disregards the class population. This
  ingredient was taken as an assumption for all the models considered
  here, which the data fully justify.  }

The model leads to the nontrivial prediction that connects the growth
exponent of a functional category to the slope of the population
family histogram restricted to the same category. In other words, the
populations functional categories and evolutionary families of genes
are connected by a simple quantitative law.
Specifically, the ratio between the exponent of the cumulative
histogram of all families and the exponent of the cumulative histogram
restricted to families belonging to a functional category is predicted
to be equal to the exponent for the scaling of the functional
category.

\begin{figure}[t]
\begin{center}
\includegraphics[width=\columnwidth]{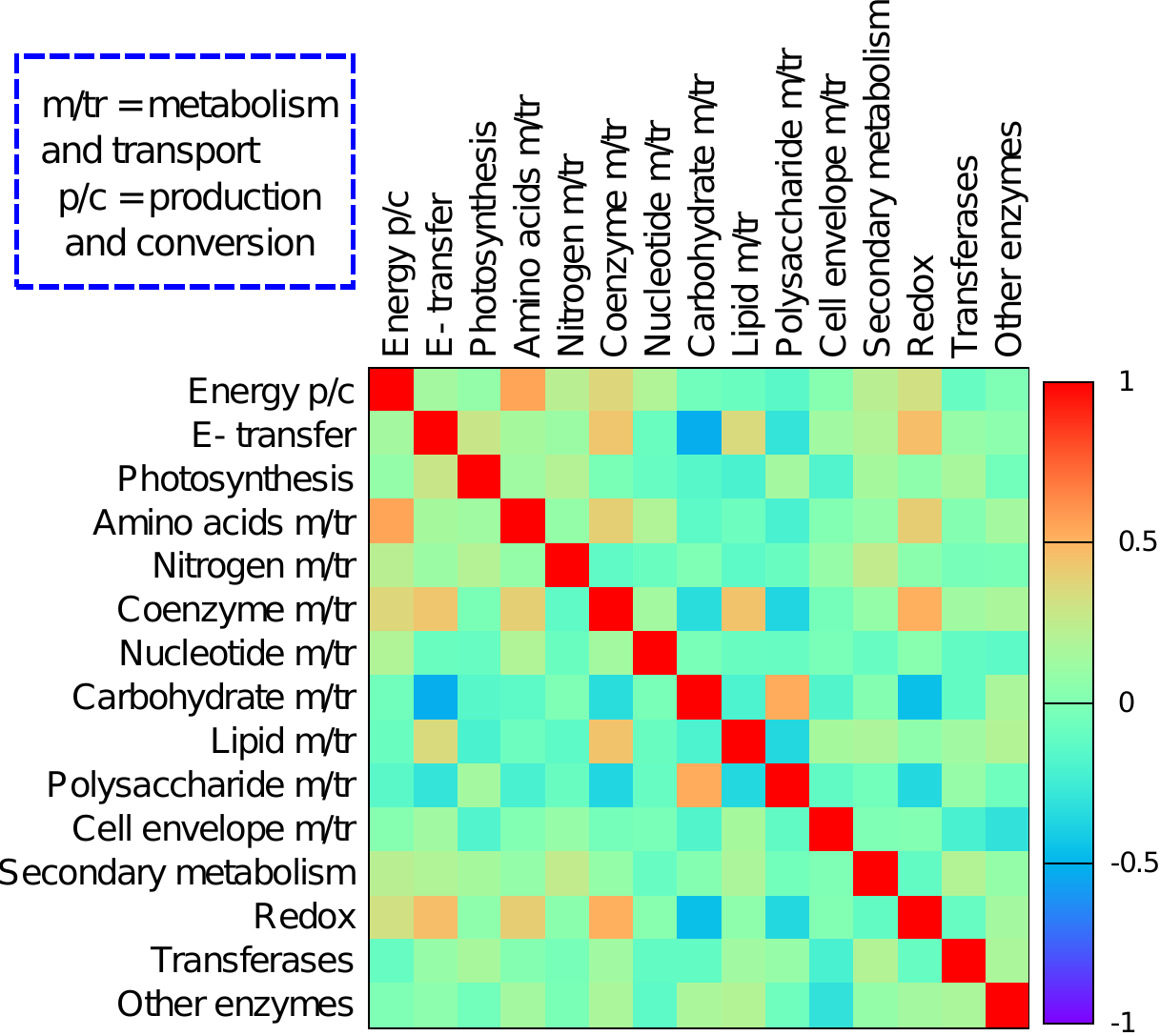}
\end{center}
\caption{ Correlation between the populations of $ 24 $ different
    metabolic functional categories from the SUPERFAMILY database for
    $ 753 $ bacteria. The correlation matrix is calculated from
  fluctuations of categories from the average trend (see Methods).
  Both correlation and anticorrelation are present between categories.
  Different metabolism categories are highly (anti-)correlated.}
  \label{fig:correlation}
\end{figure}

To generate this prediction, we have analyzed in detail the case of
transcription factors, where the exponent of the population histogram
is halved due to the quadratic scaling using mean-field calculations
and simulations, and verified that it holds in general by simulations
of both model variants.
Empirical data on transcription factors follow this behavior
remarkably well, showing population cumulative histograms of
transcription factor superfamilies decaying with halved exponents
compared to the global populations.
The fatter tails of the TF histograms might also be related to the
fact that only a few highly populated DNA-binding domain superfamilies
dominate the population of TF DNA-binding domains and determine the
scaling laws (Supplementary Text and Supplementary
Figures~\ref{fig:supp-TFclass} and~\ref{fig:supp-smalltf}).
More in general, we have also compared the behavior of domain family
population histograms for all the empirical functional categories with
the prediction, obtaining results that are in good agreement (Figure
\ref{fig:expcorr}), in particular for the highly populated categories,
where the fitting procedure gives the highest confidence.
The only highly-populated category that significantly violates
this general trend is small molecule binding, a category composed of
very few highly-populated domain families. This category is known to
follow peculiar evolutionary laws, with high mobility of domains
across the metabolic network, resulting in members of the same
family being scattered across different pathways and producing
lineage-specific domain families, with frequent re-invention of the
same function by different families~\cite{Koonin01,Chothia2003}.
Thus, the exception makes biological sense, and can be understood in
terms of members of evolutionary classes ``jumping'' to different
functional categories with high rate during evolution.

\subsection*{Correlated and absolute recipes}

The central ingredient of our main model \rev{(model I)} is the
coupling between addition/removal of genes in different functional
categories.  From a biological standpoint, it is reasonable that gene
repertoires of functional categories related to each other via shared
tasks, pathways or processes should follow coordinated
rules~\cite{Koonin}.  In order to further justify this assumption, we
probed directly the empirical domain data for correlation between
number of domains in different functional categories,. To this end,
for each genome $g$ we calculated the deviation $\delta n_c(g)$
between the functional category size ($n_c(g)$ and its average size in
genomes of comparable size (see Methods).  We then calculated the
matrix of correlations between values of $\delta n_c$ for different
functional categories $c$.
The results are reported in Figure~\ref{fig:correlation} \rev{ and
  Supplementary Tables~S3 and~S4. We also tested that this procedure
  for evaluating the correlation was not dependent on genome size
  (Supplementary Figure~\ref{fig:supp-metcorrsize}.) }
The metabolism categories appear to be highly (anti-)correlated with
each other, probably because of the role they play in different
pathways of a common metabolic network~\cite{maslov}.
The observed correlations between metabolic families might also be
relevant for reproducing the correct tail of the family population
histogram restricted to the metabolism category (Supplementary
Figure~\ref{fig:supp-comparision}).

\begin{figure*}[tp]
\begin{center}
\includegraphics[width=0.7\textwidth]{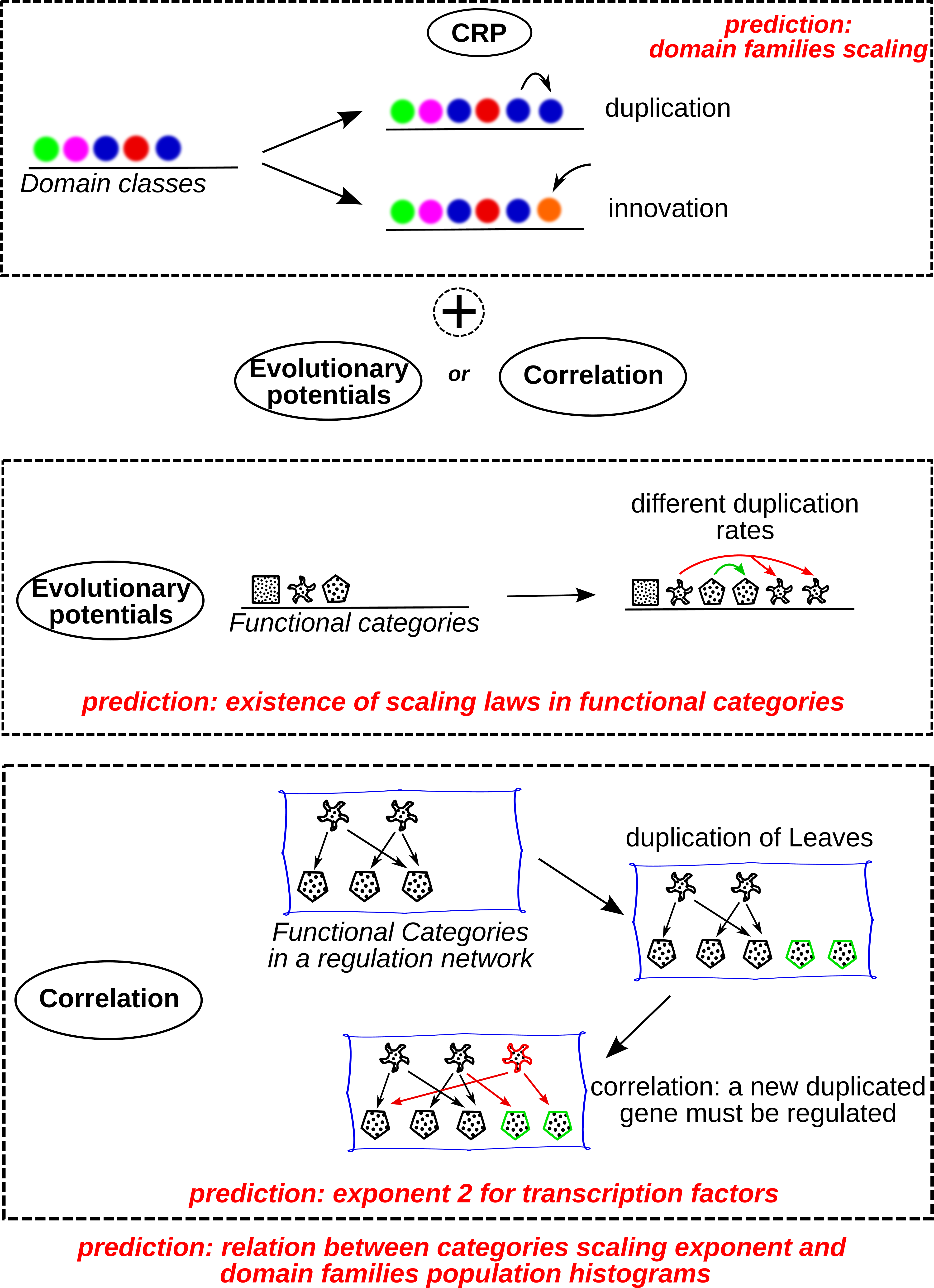}
\end{center}
\caption{ Models with correlated versus absolute moves.  Top: the
  Chinese restaurant process (CRP) acts on the homology families
  (colors) with a duplication and an innovation move. It is extended
  here to include functional categories (shapes) Middle: model with
  evolutionary potentials. Functional categories are assigned
  differential duplication rates as in ref~\cite{vannim_08}. Bottom:
  Model with correlated moves. Members of the functional categories
  are added proportionally between correlated pairs of functions
  (e.g. transcription factors and metabolic targets) as in
  Ref.~\cite{maslov}. }
\label{fig:models}
\end{figure*}

An alternative approach is a description where the growth of each
category is governed by intrinsic ``evolutionary
potentials''~\cite{vannim_08}. We have also analyzed such a
description in some detail (see Supplementary text and Supplementary
Figure~\ref{fig:supp-comparision}).
Despite of minor differences, a model \rev{combining
  class-expansion/duplication/loss with uncorrelated moves for the
  functional categories, model II,} can also perform well in reproducing the
joint scaling law and in predicting a relationship between the scaling
exponents and the functional categories. \rev{In particular, this
  means that the latter result should not by itself be considered a
  piece of evidence in favor of a correlated recipe. }
Figure \ref{fig:models} illustrates the basic differences between the
two descriptions.
The evolutionary potentials approach generically requires a lower
number of parameters, but suffers from the tedious technical problem
that the values of the growth coefficients cannot be controlled
directly, because of the scaling of the normalization constant with
genome size (see Supplementary Text \rev{and Supplementary
  Figure~\ref{fig:supp-cn}}).
The correlated model is technically more under control, since its
behavior does not rely on any unknown normalization constant.  For
this reason, it also performs better with functional categories that
grow faster than linear with genome size, such as transcription
factors.
\rev{On the other hand}, such a model can be formulated with very few
parameters \rev{only when} a synthetic description for the
correlations, such as the toolbox model, is provided.

\rev{ Here, we have considered mainly a model with three categories
  (transcription factors, metabolic, and others) and one nonzero
  correlation between metabolic domains and transcription factors.  In
  general, specific biological details of how categories are
  correlated with each other determine the scaling exponents relating
  their genome fractions to each other and genome size.
  Note that the task of formulating a correlated model for many
  categories requires a knowledge of how the different functional
  categories are ``slaved'' to each other. This structure is largely
  unknown quantitatively, and can in principle define an arbitrarily
  complex network of interactions, since many categories may correlate
  with many others in potentially complicated ways.  
  Should the importance of correlated recipes be confirmed by further
  analysis, it seems likely that the full formulation of such a
  description would still require to solve this problem.
  In order to show explicitly that the model can in principle be
  successfully extended to many categories (and still give scaling
  laws) we have analyzed the case of a simple hierarchical structure
  where many categories are slaved to a main one (see Supplementary
  Figure~\ref{fig:supp-pwrsim}). 
}

Overall, since functional categories scaling laws effectively emerge
from the correlated approach, a good reconciliation of the two
approaches could be to interpret the evolutionary potential model as
an emergent description (which can be very useful in concrete
empirical applications).
In other words, evolutionary potentials would describe emergent
effective growth of functional categories of a genome, averaging over
more ``microscopic'' evolutionary processes where addition of genes
belonging to specific functional categories needs to comply to
constraints combining different functions to perform specific cell
tasks. These kind of interactions between functions are better
described by correlated growth of functional categories.  In this
view, genome growth would be governed by a relative recipe, where the
proportions are more important than the exact amounts, rather than an
absolute recipe, where only the detailed amounts of each ingredient
play a role.

\afterpage{\clearpage}
\newpage
\bibliography{biblio}

\begin{thebibliography}{10}

\bibitem{HvN98}
Huynen, M.~A. and van Nimwegen, E. (1998)
The frequency distribution of gene family sizes in complete genomes..
{\em Mol Biol Evol,} {\bf 15}(5), 583--9.

\bibitem{vannim_03}
van Nimwegen, E. (2003)
Scaling laws in the functional content of genomes.
{\em Trends in Genetics,} {\bf 19}(9), 479--484.

\bibitem{vannim_08}
Molina, N. and van Nimwegen, E. (2008)
The evolution of domain-content in bacterial genomes.
{\em Biology Direct,} (3), 51+.

\bibitem{KWK02}
Koonin, E.~V., Wolf, Y.~I., and Karev, G.~P. (2002)
The structure of the protein universe and genome evolution..
{\em Nature,} {\bf 420}(6912), 218--23.

\bibitem{DSS02}
Dokholyan, N.~V., Shakhnovich, B., and Shakhnovich, E.~I. (2002)
Expanding protein universe and its origin from the biological {B}ig {B}ang..
{\em Proc Natl Acad Sci U S A,} {\bf 99}(22), 14132--6.

\bibitem{bassetti}
{Cosentino Lagomarsino}, M., Sellerio, A., Heijning, P., and Bassetti, B.
  (2009)
Universal features in the genome-level evolution of protein domains.
{\em Genome Biology,} {\bf 10}(1), R12+.

\bibitem{messico}
Perez-Rueda, E., Janga, S., and Martinez-Antonio, A. (2009)
Scaling relationship in the gene content of transcriptional machinery in
  bacteria..
{\em Molecular Biosystems,}
DOI: 10.1039/b907384a.

\bibitem{Koonin}
Koonin, E.~V. and Wolf, Y.~I. (2008)
{Genomics of bacteria and archaea: the emerging dynamic view of the prokaryotic
  world}.
{\em Nucleic Acids Research,} {\bf 36}(21), 6688--6719.

\bibitem{Treangen2011a}
Treangen, T.~J. and Rocha, E. P.~C. (2011)
Horizontal transfer, not duplication, drives the expansion of protein families
  in prokaryotes..
{\em PLoS Genet,} {\bf 7}(1), e1001284.

\bibitem{KWR+02}
Karev, G.~P., Wolf, Y.~I., Rzhetsky, A.~Y., Berezovskaya, F.~S., and Koonin,
  E.~V. (2002)
Birth and death of protein domains: a simple model of evolution explains power
  law behavior..
{\em BMC Evol Biol,} {\bf 2}(1), 18.

\bibitem{QLG01}
Qian, J., Luscombe, N.~M., and Gerstein, M. (2001)
Protein family and fold occurrence in genomes: power-law behaviour and
  evolutionary model..
{\em J Mol Biol,} {\bf 313}(4), 673--81.

\bibitem{KLQ+06}
Kamal, M., Luscombe, N., Qian, J., and Gerstein, M. (2006)
Analytical {E}volutionary {M}odel for {P}rotein {F}old {O}ccurrence in
  {G}enomes, {A}ccounting for the {E}ffects of {G}ene {D}uplication,
  {D}eletion, {A}cquisition and {S}elective {P}ressure.
In Koonin, E., Wolf, Y., and Karev, G., (eds.), \emph{Power {L}aws,
  {S}cale-{F}ree {N}etworks and {G}enome {B}iology},  pp. 165--193 Spinger, New
  York.

\bibitem{DS05}
Durrett, R. and Schweinsberg, J. (2005)
Power laws for family sizes in a duplication model.
{\em Ann. Probab.,} {\bf 33}(6), 2094--2126.

\bibitem{angelini}
Angelini, A., Amato, A., Bianconi, G., Bassetti, B., and Cosentino~Lagomarsino,
  M. (2010)
Mean-field methods in evolutionary duplication-innovation-loss models for the
  genome-level repertoire of protein domains.
{\em Phys. Rev. E,} {\bf 81}(2), 021919.

\bibitem{maslov}
Maslov, S., Krishna, S., Pang, T., and Sneppen, K. (2009)
Toolbox model of evolution of prokaryotic metabolic networks and their
  regulation.
{\em Proceedings of the National Academy of Sciences,} {\bf 106}(19),
  9743--9748.

\bibitem{isambert}
Isambert, H. and Stein, R. (2009)
On the need for widespread horizontal gene transfers under genome size
  constraint..
{\em Biology Direct,} {\bf 4}(1), 28+.

\bibitem{BBK+05}
Bornberg-Bauer, E., Beaussart, F., Kummerfeld, S.~K., Teichmann, S.~A., and
  Weiner, 3rd, J. (2005)
The evolution of domain arrangements in proteins and interaction networks..
{\em Cell Mol Life Sci,} {\bf 62}(4), 435--45.

\bibitem{supfam_funz}
Wilson, D., Madera, M., Vogel, C., Chothia, C., and Gough, J. (2007)
The SUPERFAMILY database in 2007: families and function.
{\em Nucleic Acids Res,} pp. D308--D313
Updated 7/2009.

\bibitem{pitman}
Pitman, J. (2006)
Combinatorial Stochastic Process,
Notes for {S}t. {F}lour {S}ummer {S}choolSpringer-Verlag, Berlin.

\bibitem{Koonin01}
Anantharaman, V., Koonin, E.~V., and Aravind, L. (2001)
Regulatory potential, phyletic distribution and evolution of ancient,
  intracellular small-molecule-binding domains..
{\em J Mol Biol,} {\bf 307}(5), 1271--1292.

\bibitem{Chothia2003}
Chothia, C., Gough, J., Vogel, C., and Teichmann, S.~A. (2003)
Evolution of the protein repertoire..
{\em Science,} {\bf 300}(5626), 1701--1703.

\bibitem{derek}
Charoensawan, V., Wilson, D., and Teichmann, S.~A. (2010)
{Genomic repertoires of DNA-binding transcription factors across the tree of
  life}.
{\em Nucleic Acids Research,}.

\end{thebibliography}

\clearpage
\newpage

\renewcommand{\thesection}{S\arabic{section}}
\setcounter{figure}{0}
\setcounter{page}{1}
\setcounter{table}{0}
\setcounter{section}{0}
\setcounter{equation}{0}

\renewcommand{\figurename}{Supplementary Figure}
\renewcommand{\tablename}{Supplementary Table}
\renewcommand{\thefigure}{S\arabic{figure}}
\renewcommand{\thetable}{S\arabic{table}}
\renewcommand{\theequation}{S\arabic{equation}}



\section*{\Large Supplementary Text and Figures for Grilli \emph{et al.}}

\vspace{2cm}



\section{Description of the model and basic mean-field results}

\rev{ This section discusses in more detail the analytical derivation
  of the scaling for the main observables of model I and II using a
  mean-field approach.}

\rev{ Consider a joint partioning of elementary units (domains or
  genes) in functional and evolutionary categories, as illustrated in
  Figure~\ref{fig:partitioning} of the main text. The elementary units
  (in our case domains), belong to a single evolutionary family $i$,
  and every family $i$ belongs to one and only one functional category
  $c$.}

\rev{The generic stochastic growth model considered here defines how
  new units are introduced into the system. The model is specified by
  a set of basic rates.  The basic set of rates is constitued by the
  probabilities $p_i$ that a newly added unit belongs to a certain
  class $i$.  More in detail, we define a probability $p_O^i$ (where
  $O$ stands for ``old'') that a new domain belongs to a family $i$
  which is already present in the system (i.e. having at least one
  member) and the probability $p_N$ (where $N$ stands for ``new'')
  that the added unit belongs to a family which is not already present
  in the system. }

\rev{The choice of $p_O^i$ and $p_N$ defines the model as a stochastic
  process for the basic observables (such as genome size $n$, family
  number $f$ and its population $n_i$, etc.), but one extra detail is
  needed. When a new class is introduced, the model needs to specify
  the category it belongs to. As discussed in the main text, in the
  model considered here a newly added family always belongs to a
  category $c$ with probability $\chi_c$.  The probabilities $p_O^i$,
  $p_N$ and $\chi_c$ can depend, in principle, on the number of units
  $n$ and on their distribution in families, on the total number of
  families $f$ and so on. Empirical data indicate (see
  Figure~\ref{fig:Chi_c} in the main text) that $\chi_c$ is a
  category-dependent constant, and thus does not depend on $n$.}

\rev{The mean-field approximation is useful to extract the basic
  information from the model~\cite{bassetti}. In each realization of
  the full stochastic process, the probabilities of the possible
  configurations at time $t+1$ are determined by the configuration at
  time $t$. The mean-field approximation assumes that the
  configuration at time $t$ is the average configuration.  For
  example, if one is interested in the number of domains belonging to
  family $i$, the average number of elements $n_i(t+1)$ at time $t+1$
  will be equal to the average number of elements $n_i(t)$ at time $t$
  summed with the average number of elements added in a time step,
  i.e. $p_O^i$. For asymptotically large $t$ this implies the
  approximate equation $\partial_t n_i = p_O^i$ for the averages (here
  the averaging procedure is implicit in the notation). Since
  typically, at each step one and only one element is added, the mean
  number of elements is $n=t$. If this is not the case, we can obtain
  $\partial_n n_i$ simply from $\partial_t n_i$ divided by $\partial_t
  n$.  Considering $n=t$ we obtain, for a generic model, the following
  mean-field equations
\begin{equation}
\displaystyle
\begin{split}
& \partial_n n_i = p_O^i \\
& \partial_n f = p_N \\
& \partial_n f_c = \chi_c p_N \\
& \partial_n n_c = \partial_n \sum_{i \in c} n_i = \sum_{i \in c} \partial_n n_i + \partial_n f_c = \sum_{i \in c} p_O^i + \chi_c p_N \ .
\end{split}
\label{eq:meanpart}
\end{equation}
}

\subsection{\rev{ Models with correlations}}
\label{sec:supp-modI}

\rev{We now deal with the scaling of the basic observables in the
  model taking into account the correlation between categories growth
  (model I of the main text).}

\rev{The correlation appears in the growth of the domain families of
  different categories. Thus the probability $p_O^i$ that a domain is
  added to a given family $i$ can be written as
\begin{equation}
  p_O^i =
  \frac{\sum_{j=1}^f a_{i,j} n_j-\alpha}
       {\sum^f_{i,j=1}  a_{i,j} n_j + \theta} \ .
\label{eq:coor}
\end{equation}
The coordinated growth of functional categories is encoded by the
coefficients $a_{i,j}$, responsible for the correlated expansion of
evolutionary families $i$ and $j$ (See Equation~\ref{eq:CRPdupl-class}
of the main text).  The standard Chinese Restaurant Process (CRP) is
obtained by imposing $a_{i,j}=\delta_{i,j}$ (where $\delta_{i,j}$ is
equal to $1$ if $i=j$ and $0$ otherwise).  We assume that these
coefficients depend only on the functional categories $c$ and $c'$ to
which the families $i$ and $j$ belong.  The probability of introducing
a new domain is given by
\begin{equation}
   p_N = \frac{\alpha f + \theta }
   {\sum^f_{i,j=1}  a_{i,j} n_j + \theta } \   .
\label{eq:coor2}
\end{equation}
}

\subsubsection{\rev{Model Ia.}}
\label{sec:supp-modIa}

\rev{We consider a model inspired by ref.~\cite{maslov} (the toolbox
  model, in which the growth of the number of transcription factors is
  coupled to the number of added metabolic enzymes), extended to
  describe a joint partitioning in functional and evolutionary
  categories.  In the original version of the model the average
  increment of the main observables at each time step is
\begin{equation}
\begin{cases}
\displaystyle
\Delta n_{met} = \frac{U}{n_{met}} \\
\Delta n_{TF} = 1 \ ,
\end{cases}
\label{eq:toolbox}
\end{equation}
and thus $\Delta n_{TF} / \Delta n_{met} = n_{met}/U$, which gives a
quadratic scaling for $n_{TF}$ with $n_{met}$.}

\rev{Model Ia is an extension of the toolbox model is formulated
  following equation~\ref{eq:coor}, by using a proper definition of
  $a_{i,j}$, such as the same equation of the toolbox model is
  valid. We observe that, for our purpose, the time step of
  equation~\label{eq:toolbox} we can be defined arbitrarily, as genome
  growth is eventually parameterized by $n$. Rewriting the equations
  as
\begin{equation}
\begin{cases}
  \displaystyle
  \Delta n_{met} = n_{met} \\
  \Delta n_{TF} = n_{met} \frac{n_{met}}{U} \ ,
\end{cases}
\label{eq:toolbox2}
\end{equation}
gives the summed probabilities $p_O^i$ relative to the two categories
\begin{equation}
\begin{cases}
  \displaystyle
  p_O^{met}:=\sum_{i \in met} p_O^i = \frac{n_{met}-\alpha f_{met}}{C(n)} \\
  p_O^{TF}:=\sum_{i \in TF} p_O^i = \frac{\frac{n_{met}}{U}
    n_{met}-\alpha f_{TF}}{C(n)} \ ,
\end{cases}
\label{eq:toolbox3}
\end{equation}
while
\begin{equation}
   p_N = \frac{\alpha f + \theta }{C(n)}  \ .
\label{eq:coor3}
\end{equation}
}

\rev{Accordingly, we extend the model to an arbitrary number of
  families by the choice $a_{i,j}=\frac{n_{met}}{U}\frac{n_i}{n_{TF}}$
  if $i$ is a $TF$ family and $j$ a metabolic family and zero
  otherwise. This gives
\begin{equation}
\begin{cases}
  \displaystyle p_O^i =
  \frac{ \sum_{j \in met} \frac{n_{met}}{U}\frac{n_i}{n_{TF}} n_j-\alpha}
       {\sum^f_{i,j=1}  a_{i,j} n_j + \theta} & \text{if $i \in TF$} \\
  \displaystyle 
   p_O^i = \frac{n_i -\alpha}
                {\sum^f_{i,j=1} a_{i,j} n_j + \theta}
  & \text{if $i \in met$ \ .} \\
\end{cases}
\label{eq:defIa}
\end{equation}
} 

\rev{ This model gives the asymptotic quadratic scaling of $n_{TF}$
  with $n_{met}$ by definition, using the exact same argument as the
  toolbox model. Other results have been obtained numerically (see
  Supplementary Figure~\ref{fig:supp-ToolPan}).}

\subsubsection{\rev{Model Ib.}}
\label{sec:supp-modIb}

\rev{ This second formulation of a model with correlated recipe (model
  Ib) imposes a different correlation rule. For example, consider the
  model involving only two functional categories, transcription
  factors controlling metabolic processes and metabolic enzymes.  }

\rev{In this variant the coefficients $a_{i,j}$ have both a diagonal
  and a non diagonal part, $a_{i,j}=\delta_{i,j}+b_{i,j}$. If $b=0$
  the model is the standard Chinese Restaurant Process.  For this
  reason, model Ib is simpler to treat analytically, exploiting
  previous results.  This work focuses mainly on the case
  $b_{i,j}=n_i/n_{met}$ if $i$ is a family from the functional
  category of transcription factors and $j$ is a family from the
  metabolic functional category (and $b_{i,j}=0$ otherwise).}

\rev{In this case, the summed probabilities $p_O^i$ relative to the
  two categories are
\begin{equation}
\begin{cases}
  \displaystyle p_O^i =
  \frac{n_i + \sum_{j \in met} \frac{n_i}{n_{met}}-\alpha }
       {\sum^f_{i,j=1}  a_{i,j} n_j + \theta} & \text{if $i \in TF$} \\
  \displaystyle p_O^i =
  \frac{n_i  -\alpha}{\sum^f_{i,j=1}  a_{i,j} n_j + \theta} & \text{if $i \in met$.} \\
\end{cases}
\label{eq:defIb}
\end{equation}
Using the definitions given in Equation~\ref{eq:meanpart}, one can see
that,
\begin{equation}
  \displaystyle
  C(n)  \partial_n n_{TF} =   n_{TF} + n_{TF} 
                           - \alpha f_{TF}   + C(n) \partial_n f_{TF} =
2 n_{TF} - \alpha f_{TF} + \alpha f_{TF} + \theta \chi_{TF} = 2 n_{TF} + \theta \chi_{TF} \ ,
  \label{eq:scaling}
\end{equation}
while
\begin{equation}
  \displaystyle
  C(n)  \partial_n n_{met} =   n_{met}  
                           - \alpha f_{met}   + C(n) \partial_n f_{met} = n_{met} + \theta \chi_{met}  \ .
  \label{eq:scaling2}
\end{equation}
Hence, for large $n$, since $ \partial_n f_{c} = \chi_{c} p_N \simeq
\alpha f_{c}$, the  terms in the r.h.s. of
Equations~(\ref{eq:scaling}) and (\ref{eq:scaling2}) cancel, giving
the effective equation,
\begin{equation}
  \displaystyle
  \frac{d n_{TF}}{d n_{met}} \simeq \frac{2 n_{TF}}{n_{met}} \ ,
\end{equation}
and thus the scaling $n_{TF} \sim n_{met}^2$. }

\subsection{\rev{Model II (model with evolutionary potentials)} }
\label{sec:supp-nimwegen}

This section \rev{presents in more detail the uncorrelated version
  of the} model \rev{for the joint scaling (model II)}, assigning
evolutionary potentials~\cite{vannim_08} $ \rho_c $ to the functional
categories, related to the probability that a gene added in a
functional category is fixed by natural selection. \rev{This model is
  an example of an ``absolute recipe'', since each category grows with
  an intrisic rate $\rho_c$, summing up the growth of the families
  belonging to the given category.}
The rate $ \rho_c $ acts on family growth through the
\rev{class-expansion} move. \rev{The probability of class expansion of
  a family belonging to the category $c$ is equal to
\begin{equation}
   p_O^i =
  \frac{\rho_{c(i)} n_i-\alpha}{\sum^f_{j=1}\rho_{c(j)} n_j + \theta},
\label{eq:evp1}
\end{equation}
where $ \rho_{c(i)} = \rho_c $ if the evolutionary family $ i $
belongs to the functional category $ c$. This model assumes that the
value of $\rho_c(i)$ depends only on the category to which family $i$
belongs. The probability that a domain belonging to category $c$ is
added by class expansion is then
\begin{equation}
   p_O^c := \sum_{i \in c} p_O^i =
  \frac{\rho_{c} n_c-\alpha f_c }{\sum^f_{j=1}\rho_{c(j)} n_j + \theta} .
\label{eq:evp1b}
\end{equation}
Equally, the probability that the new domain is introduced by an
innovation move (i.e. it belongs to a new family) is equal to
\begin{equation}
p_N = \frac{\alpha f +
  \theta}{\sum^f_{j=1}\rho_{c(j)} n_j + \theta}.
\label{eq:evp2}
\end{equation}
Under the assumption (confirmed by empirical data, \rev{ see main
  text}) that the growth of old functional categories by adding new
homology families through the innovation move is uniform (i.e. that $
f_c = A_c + \chi_c f $), the probability that a new family belonging
to the category $c$ is added by an innovation move is
\begin{equation}
p_N^c := \chi_c p_N = \chi_c \frac{\alpha f +
  \theta}{\sum^f_{j=1}\rho_{c(j)} n_j + \theta} = \frac{\alpha f_c +
  \theta \chi_c}{\sum^f_{j=1}\rho_{c(j)} n_j + \theta}.
\label{eq:evp2b}
\end{equation}
}

\subsection*{Evolutionary potentials can reproduce the combined scaling
  laws at finite sizes.}

We tested this model by a combination of mean-field analytical
arguments and direct simulation.

\rev{The mean-field equations are obtained from
  Equation~\ref{eq:meanpart} by using Equations~\ref{eq:evp1}
  and~\ref{eq:evp2}.  The equation for the growth of the mean number
  of members $n_c$ of a functional category can be obtained simply by
  summing on the homology families that belong to a given category,}
\begin{equation}
  \partial_n n_c = \frac{ \rho_c n_c + \theta \chi_c }{C(n)} \ ,
\label{eq:catgrow1}
\end{equation}
where $ C(n) \simeq \sum_i \rho_i n_i $. If $C(n) \sim n$, equation
(\ref{eq:catgrow1}) corresponds to the evolution equation written by
Molina and Nimwegen.  Simulations of this model (see Supplementary
Figure~\ref{fig:supp-cn}) confirm that this is the case. Thus, the
mean-field argument predicts that this model can reproduce both
scaling laws.

Also note that a rescaling of $C(n)$ is equivalent to a rescaling of
$\alpha$.  Indeed, for large $n$, $p_N \simeq \alpha f / C(n) $ (and
$p_O = 1- p_N$), so imposing $C(n) \simeq q n$ is equivalent to
dividing $\alpha$ by the constant factor $q$. Thus, one can choose $q
= 1$ without loss of generality (by a rescaling of all the $ \rho_c
$), and the solution for the population of a functional category will
be $n_c \sim n^{\rho_c/q}$ as in the Molina/Nimwegen model, \rev{and
  thus} $ \zeta_c = \rho_c / q $

On the other hand, an important point regarding this model is that,
asymptotically for any choice over the $ \rho_c $ set, the maximum
large-$n$ exponent observed will be $ 1 $, Indeed, we can use the
approximation $ C(n) = \sum_i \rho_i N_i \sim \rho_{\mathrm{max}}
n^{\rho_c/q} $, but $ C = q n $, so that $ q = \rho_{c_{max}}$. This
means that an exponent close to $2$, such as that observed for
transcription factors can only be obtained in a transient regime of
the model.
Furthermore, the change of the evolutionary potential of one
functional category has repercussions on the other categories, as it
implies a change in the normalization costant $C$.  These facts make a
direct identification of the value of the evolutionary potential with
an intrinsic propery of a single functional category difficult. They
also make the direct identification of evolutionary potentials less
straightforward (as it requires an arbitrary rescaling).

However, the above remarks have little practical importance, and the
large-$n$ behaviour of the model does not really affect its performance
at the relevant values of $n$.
Numerical simulations show that at the empirical genome sizes, the
scaling behaviour of the model can reproduce rather well the empirical
one.  For simplicity we have restricted to three main categories
(transcription factors, metabolic genes and ``others'') and we
verified that in practice it is not hard to find a parameter set in
good agreement with the empirical data on protein domains
(Supplementary Figure~\ref{fig:supp-comparision}). The general number
of parameters to adjust increases with the number of functional
categories that one needs to consider.

\section{Exponents of family size distribution histograms}
\label{sec:supp-histoexp}

This section discusses the family size distribution histograms, as
obtained from the mean-field approach.  \rev{To fix the ideas, we will
  focus on model Ib, where the mean-field equations can exploit the
  known results from the CRP. } It is possible to write a mean-field
``flux equation'' for the histograms~\cite{angelini}, which implements
the fact that each duplication adds a family with one extra member to
the histogram count and subtracts a family with its previous
population,
\begin{equation}
  \partial_n f(d,n) = p_O(d-1,n) f(d-1,n) - p_O(d,n) f(d,n) + p_N \delta_{d,1}
\label{eq:flux}
\end{equation}
where $ p_O(d,n) = \frac{d-\alpha}{n+\theta} $ is the probability that
a family with $ d $ domains add a new duplicated member. \rev{The term
  $p_N = \frac{\alpha f + \theta}{n + \theta}$ contains the innovation
  probability contributing to the growth of the number of families
  with one member.} Note that the flow between families can be written
as
\begin{equation*}
  \displaystyle
  \sum_{
          i \in \left\{
                   \stackrel {\mathrm{families \   with}}
                             {j \mathrm{\   domains}}   \right\}
        }
\partial_n n_i = (d-\alpha) \frac{f(d,n)}{n+\theta}.
\end{equation*}
This equation requires an assumption on $ f(d,n)$ in order to be
solved.  
\rev{We assume the ansatz $ f(d,n) = P(d) f(n) $ which is justified by
  both simulation and empirical data~\cite{angelini}}. Using the fact
that $ \partial_n f(n) = p_N$, combined with Equation~\ref{eq:flux}
gives the following equation for the probability of a family to have
$d$ members
\begin{equation}
  \alpha P(d) = (d-1 - \alpha) P(d-1) - (d-\alpha) P(d) \ \ ,
\end{equation}
which can be solved in discrete or continuous $ d $ to get
\begin{equation}
\displaystyle
P(d) \sim \left( \frac{1}{d} \right)^{1+\alpha} \ .
\label{eq:HistoTot}
\end{equation}
This predicts the asymptotic behaviour of data and simulations (see
Figure~\ref{fig:expcorr}) with $ \beta = \alpha $, where $\beta$ is
the asymptotic exponent of the family size distribution.

\rev{Let us now turn to the same distribution, restricted to
  transcription factors.  } In \rev{model Ib}, the flux from
transcription factor families caused by family expansion is caused by
two separate contribution, the CRP standard one, plus \rev{additions
  of transcription factors to an existing family caused by the addition of
  a metabolic enzyme}
\begin{equation}
  \displaystyle
  p_O^{i}(n) = \frac{1}{C(n)} \left[ (n_{i} -\alpha) +
    \frac{n_{i}}{n_{met}} n_{met}   \right] \text{, if $i \in TF$}
\end{equation}
i.e.
\begin{equation}
\displaystyle
   p_O^{i}(n) = \frac{1}{C(n)} \left[ 2 n_{i}   -\alpha  \right] \text{, if $i \in TF$.}
\end{equation}
\rev{ Thus, for the transcription factor families, the probability
  that a domain is added to a family with $d$ members will be
\begin{equation}
\displaystyle
   p_O^{TF}(d,n) = \frac{1}{C(n)} \left[ 2 d   -\alpha  \right] \ .
\label{eq:p_oTF}
\end{equation}
The quantity $p_O^{TF}(d,n)$ is the probability that a new
transcription factor domain is added to a family with $d$ members.  }
The flux equation for TF families \rev{can be obtained by substituting
  equation~\ref{eq:p_oTF} in equation~\ref{eq:flux}, } 
(for $d>1$)
\begin{equation}
  \displaystyle
  C(n) \partial_n f_{TF}(d,n) = \left[ 2(d-1) -\alpha  \right] f_{TF}(d-1,n) 
  -  \left[ 2 d -\alpha  \right] f_{TF}(d,n)
\end{equation}
This is solved using the usual ansatz $f_{TF}(d,n) = P_{TF}(d)
f_{TF}(n)$ \rev{(as explained above it is confirmed by both data and
  simulations)}.  Using $ f_{TF}(n) = \chi_{TF} f(n) $,
leads to the equation
\begin{equation}
  \alpha P_{TF}(d) = (2d-2 - \alpha) P_{TF}(d) - (2d-\alpha) P_{TF}(d) \ \ ,
\label{eq:PPTF}
\end{equation}
which gives:
\begin{equation}
  \displaystyle
  P(d)_{TF} \sim \left( \frac{1}{d} \right)^{1+\frac{\alpha}{2}} \ \ ,
\label{eq:HistoTF}
\end{equation}
that is $ \beta = \alpha / 2 = \beta / 2 $.  In the above calculation
we have supposed again that the number of transcription factors is
small with respect to to the total number of metabolic enzymes.

Furthermore, it can be argued that this fact is more general. Indeed,
each time the per-family duplication probability for the TF functional
category will have the form
\begin{displaymath}
  p_{O}^i \simeq 2 n_i \ ,
\end{displaymath}
when family $ i $ belongs to TF category, the coefficient $2$ will
appear in the equation for $P(d)_{TF}$ modifying the exponent.  In
particular, this will also be true for \rev{models Ia (generalizing the
toolbox model) and II (generalizing evolutionary potentials)}.

In other words, each time a functional category scales with a given
exponent, it can be argued on rather general grounds that the exponent
of the population histograms of the homology families that form it will
be affected.
It is possible to to generalize this argument, and find a precise
relationship between the scaling exponent of a category and the family
population histogram (restricted to the same category). In other
words, if $ \zeta_c $ is the scaling exponent of the category $ c $
and $ \beta_c $ is the exponent of the cumulative distribution
histogram for the families belonging to category $c$, that is (see
Equation~\ref{eq:HistoTF}):
\begin{displaymath}
    P(d)_c \sim \left( \frac{1}{d} \right)^{1+\beta_c} \ \ ,
\end{displaymath}
we suggest that \rev{$ \beta_c = \beta / \zeta_c $}.  We tested this
prediction in empirical data plotting $ 1/\beta_c $ versus $ \rho_c $
in Figure~\ref{fig:expcorr} (Pearson correlation coefficient $ 0.47
$).

\section{Comparison of models by numerical simulation}
\label{sec:supp-comparision}

\subsection{\rev{Correlated and absolute recipes}}

This section compares the correlated duplication and the evolutionary
potential model variants.  We considered a three categories model (TF,
Metabolic and ``other'').

The evolutionary potential model needs to supply three parameters $
\rho_c $, while the correlated model needs to supply the correlation
law between categories ($ a_{ij} $).  We impose a correlation only
between transcription factor and metabolic families with the
\rev{correlated} model \rev{Ib} prescription, i.e.
\begin{equation}
  \displaystyle
a_{ij} = n_i/n_{met},
\end{equation}
where $ i $ is a TF family and $ j $ Metabolic, $ a_{ij} = 0 $ (no
correlation) otherwise.

Figure~\ref{fig:supp-comparision} summarizes the results of this
comparison.  The correlated duplication model performs better in
reproducing the behavior of the transcription factor category (both
scaling law and histograms).  Both models are unsatisfactory in
reproducing the family population histogram of the metabolism
families. This is probably caused by the fact that neither model
include a correlation between metabolic families (Figure
\ref{fig:correlation}).

\begin{figure}[tbp]
\begin{center}
\includegraphics[width=0.9\columnwidth]{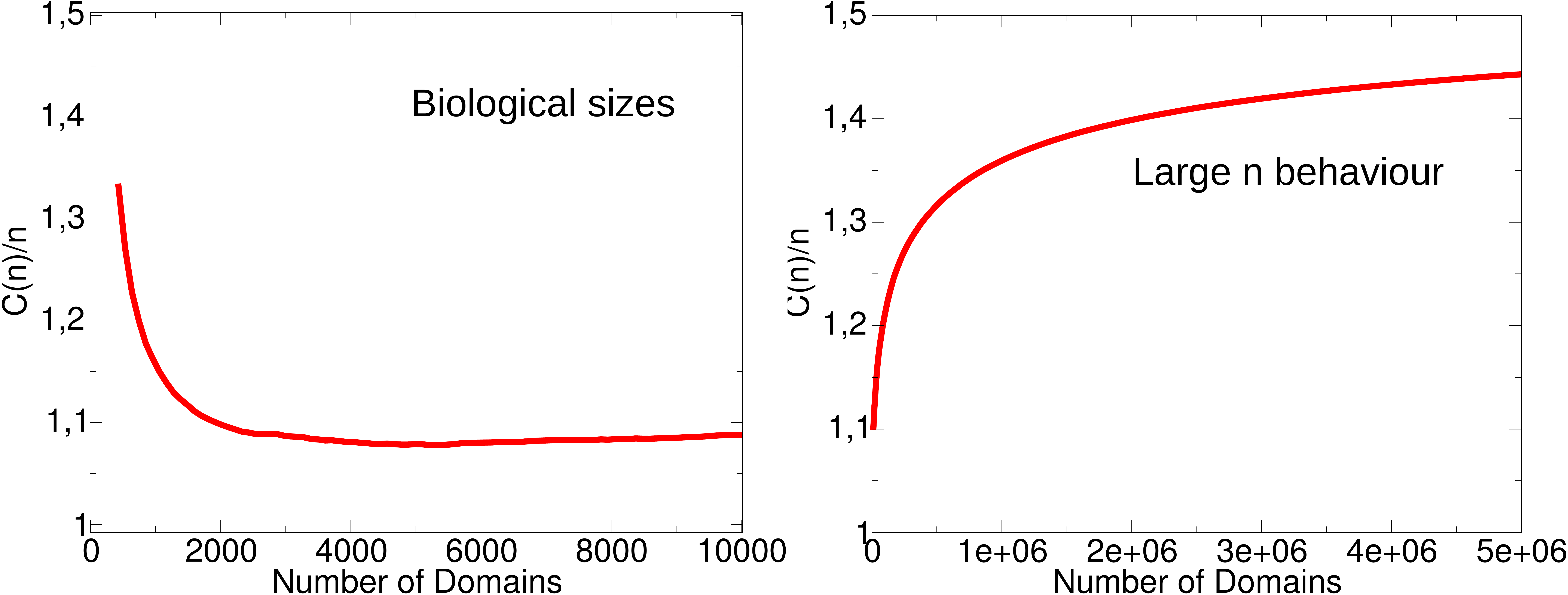}
\end{center}
\caption{ { \bf Normalization constant inthe model with evolutionary
    potentials (model II).}  Behavior of the ratio $ C(n)/n $, where $
  C(n) $ is the normalization factor for the evolutionary potential
  model. Data from simulations with three categories run at parameters
  $ \alpha = 0.3 $ and $ \theta = 140 $.  $ C(n) $ is linear with $ n
  $ in the range of empirical genome sizes, it then looses linearity,
  to become linear only asymptotically.}
  \label{fig:supp-cn}
\end{figure}

 Figure~\ref{fig:supp-cn} illustrates the behaviour of the
 normalization function $ C(n) $.  $ C(n) $ is linear with $ n $ in
 the range of empirical genome sizes (although the slope is not
 exactly $ 1 $). It becomes nonlinear at larger sizes, and its linear
 behavior is restored only at very large values of $n$.

\subsection{\rev{Model I can reproduce a set of different exponents}}

\rev{Extending a model (with absolute or correlated recipe) to a large
  number of categories is not a simple task.
  In the case of an absolute recipe model, adding a new category $c'$
  (and thus introducing a new evolutionary potential $\rho_{c'}$)
  generally requires, in order preserve the scaling of all the
  categories, a tuning of all the evolutionary potentials (both the
  old ones and the new one). This is due to the fact that all the
  evolutionary potentials appear in the normalization constant $C(n)$
  in the growth equation of each category (Equation~(\ref{eq:evp1}).
  In a model with a correlated recipe, the main problem is related to
  the fact that the interaction laws between categories are not known,
  they can be complex and possibly include feedback.}

\rev{In order to produce the proof of principle that a model with
  correlated recipes can work with more than three categories, we
  considered a trivial generalization of model Ib to multiple
  categories that are slaved to a main one, and considered the
  question of whether this model would be able to reproduce an
  arbitrary set of scaling exponents for the categories.}

\begin{figure}[tbp]
\begin{center}
\includegraphics[width=0.9\columnwidth]{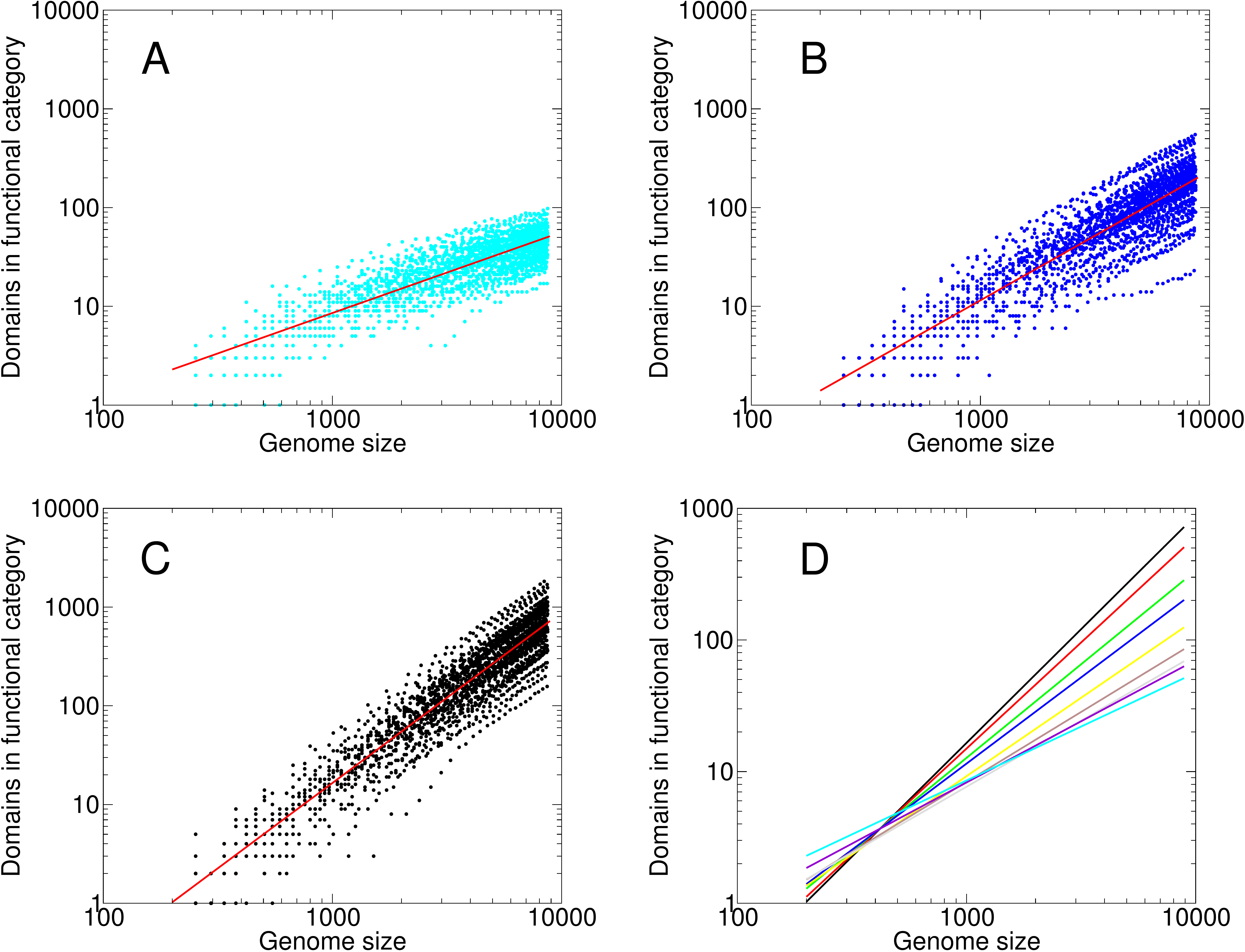}
\end{center}
\caption{ \rev{{ \bf Simulation of model Ib with $10+1$ categories.}
    $10$ categories are slaved to one master category with different
    correlation laws, which determine the observed exponents).  Panel
    A, B and C show the simulations of the population of three
    categories (respectively with $\gamma_c$ equal to $1$, $0$ and
    $-0.7$).  The red lines are power-law fits of the simulated
    data. Panel D shows the power-law fits of the simulated data for
    all ten categories.}}
  \label{fig:supp-pwrsim}
\end{figure}

\rev{We consider a correlation matrix $a_{i,j}$ of the form
  $\delta_{i,j}+b_{i,j}$, where $b_{i,i}=0$.  This model deals with
  $\cal{C}$$+ 1$  categories, the $met$ category (in analogy with model Ib
  defined in the main text, this is a category whose growth is not
  conditioned to the others), and an additional set of $\cal{C}$
  categories labeled from $1$ to $\cal{C}$. The non diagonal
  correlation coefficients $b_{i,j}$ are zero if family $i$ belongs to
  the $met$ category, and $\gamma_{c(i)} n_i / n_{met}$ if family $i$
  belongs to category $c$, different from $met$, and $j$ belongs to the
  $met$ category. Substituting this choice in equation~\ref{eq:defIb},
  gives
\begin{equation}
\displaystyle
\frac{d n_c}{d n_{met}} = (1+\gamma_c) \frac{n_c}{n_{met} }
\end{equation}
and thus
\begin{equation}
\displaystyle
n_c \sim n_{met}^{1+\gamma_c} \ .
\label{eq:predgamma}
\end{equation}
Supplementary Figure~\ref{fig:supp-pwrsim} shows simulations from a
model with $10+1$ categories.  The model is able to reproduce an
arbitrary set of exponents. We observe that this version has similar
problems as the model with evolutionary potentials, as, in absence of
a biological underlying model, it needs the tuning of a set of
parameters to reproduce the scaling laws.}
\rev{The fitted exponent is typically different from $1+\gamma_c$,
  specifically it seems to be closer to one. We interpret this as a
  finite size effect, due to the fact that the contribution of
  innovation to the scaling exponents is relevant.}


\section{Details of TF-domain superfamily scaling}
\label{sec:supp-TFclasses}

We observe that the quadratic (or very nearly so) scaling for
transcription factors is clearly visible at in the two most populated
families of transcription factor DNA-binding domains (Homeodomain-like
and Winged-helix), which have a rather clean slope (see Supplementary
Figure~\ref{fig:supp-TFclass}). In fact, three families present a
clearly observable scaling alone (Homeodomain-like, Winged-helix and
C-terminal), but just the first two follow a very nearly quadratic
scaling.

Note however that removing the six most populated TF families,
representing $80\%$ of the total TF-domain population, the remaining
ones considered together still present a scaling when added up, but
with exponent \rev{$\simeq 0.9 $ (see Supplementary
  Figure~\ref{fig:supp-smalltf})}.
This indicates that the collective scaling of TF families cannot be
entirely recunducted to properties of the most populated ones, but
these are the families responsible for the \emph{quadratic} scaling.

\rev{ Thus, the ``pure'' quadratic scaling is observable in the
  largest transcription factor families. Collecting all the families,
  wemeasure a lower exponent in empirical data (close to
  $1.6$). Supplementary Figure~\ref{fig:supp-smalltf} explains this
  behavior, showing the total contribution of the smaller
  transcription factor families.  These families collectively show a
  lower exponent (close to $1$). Thus, we can interpret the lower
  collective exponent as an effect of family size (i.e., in the
  language of statistical mechanics, a ``finite-size'' effect)
  connected to the fact that for smaller family size, the innovation
  move is more relevant and thus the family expansion process is
  slower.  The same effect is present in our simulations (see
  Supplementary Figure~\ref{fig:supp-ExpSim}.)}

\clearpage

\newpage



\begin{figure}[tbp]
\begin{center}
\includegraphics[width=0.6\columnwidth]{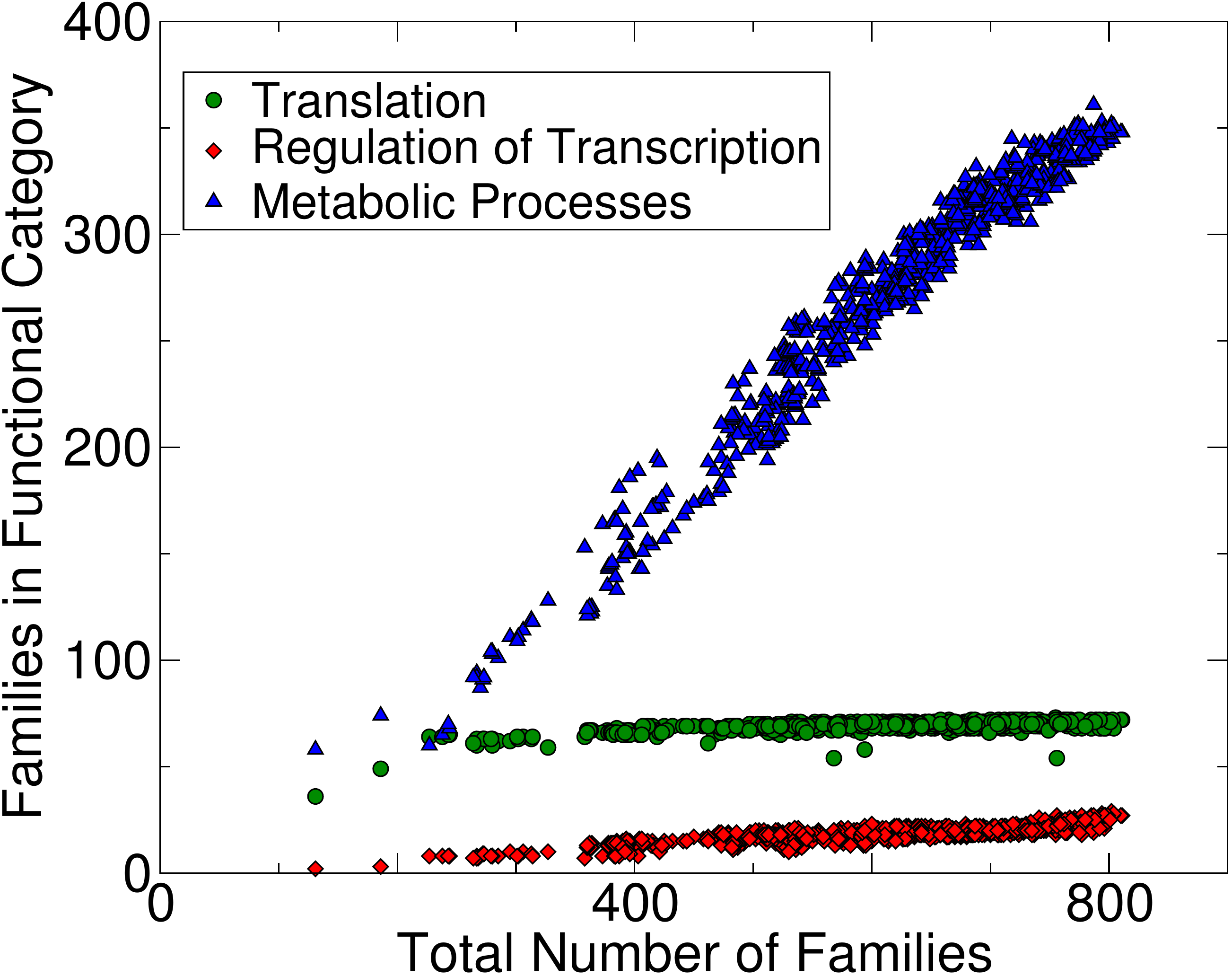}
\end{center}
\caption{ { \bf Scaling of the number of families in the three main
    functional categories.} Linear scaling behaviour of the number of
  families in three important functional categories versus total
  number of families from empirical data (for $ 753 $ bacteria in the
  SUPERFAMILY database). The slopes for the three linear laws are $
  0.01 $ (Translation), $ 0.03 $ (Regulation of Transcription) and $
  0.47 $ (Metabolic Processes).}
  \label{fig:supp-vannimSF}
\end{figure}

\begin{figure}[tbp]
\begin{center}
\includegraphics[width=0.6\columnwidth]{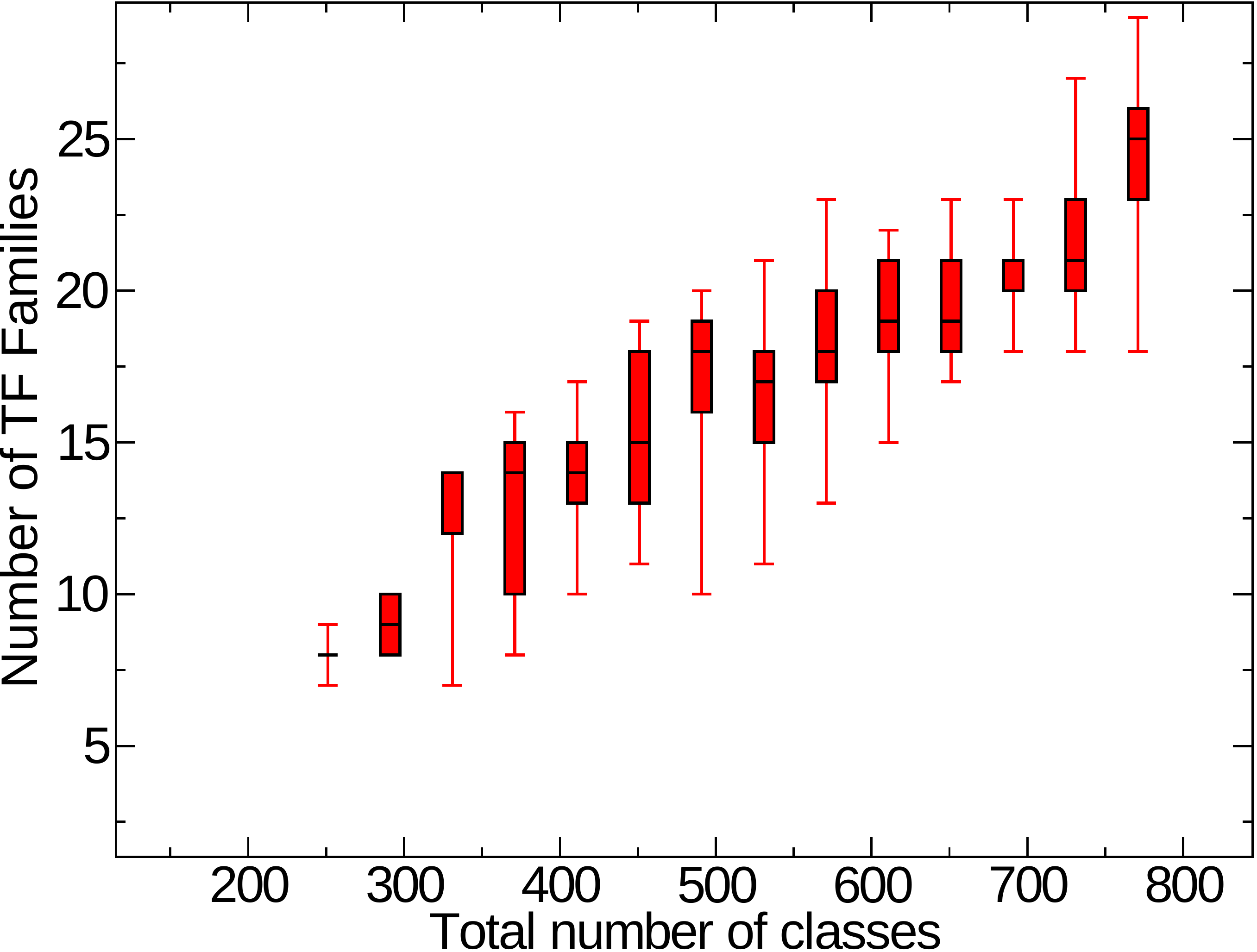}
\end{center}
\caption{ { \bf Transcription factor families.}  Boxplot of the
  number of transcription factor domain families versus total number
  of domain families (data from 753 SUPERFAMILY bacteria). There
  appears to be a roughly linear scaling. This means that the number
  of TF domain families is compatible with a null hypothesis of
  independent addition model. Charoensawan \emph{et al}~\cite{derek}
  propose that the number of TF families follows a linear scaling with
  genome \emph{size}. If this were to be the case, the innovation
  dynamics of transcription factor families should be distinct form
  other families. In fact, if $ f_{TF}(n)\sim n $, since the total
  number of families is sublinear, $ f(n) \sim n^{\alpha} $ in the CRP
  (Figure~\ref{fig:partitioning}), then one would have $
  f_{TF}~f^{2-\alpha} $, which is not confirmed by the SUPERFAMILY
  data analyzed here.  }
  \label{fig:supp-bpTFsupfam}
\end{figure}

\begin{figure}[tbp]
\begin{center}
\includegraphics[width=0.8\columnwidth]{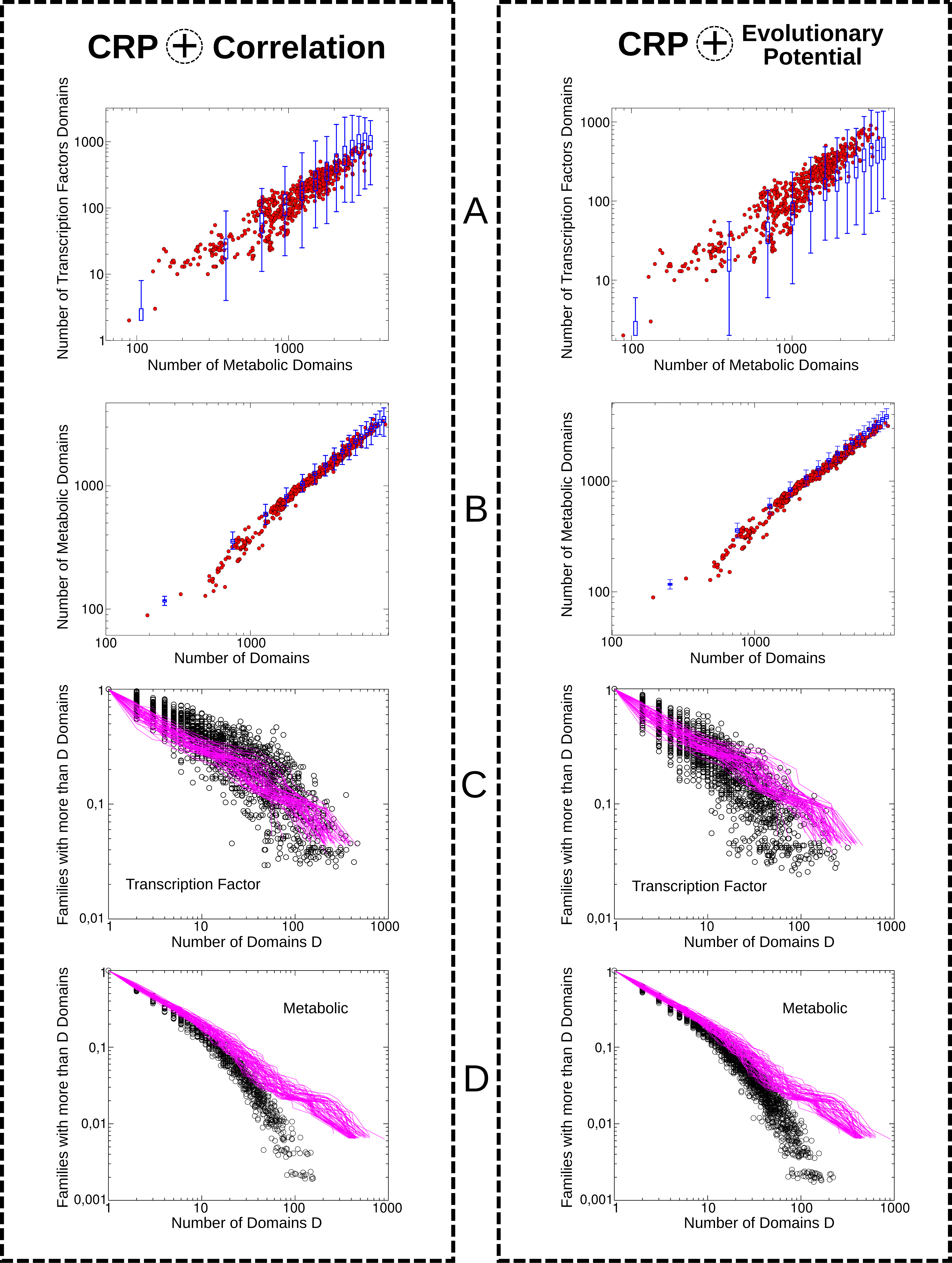}
\end{center}
\caption{ { \bf Comparison between models Ib and II.}  Comparison
  between simulation of the correlated duplication model Ib (left
  panel) and evolutionary potentials (right panel) model variants with
  empirical data.  Simulations are run at $ \alpha=0.3 $ and $ \theta =
  140 $.  (a) Number of TFs domains vs. number of metabolic domains
  (the blue boxplot corresponds to simulations, red circles to
  empirical data).  (b) Number of metabolic domains vs. total number
  of domains (the blue boxplot corresponds to simulations, red
  circles to empirical data).  (c) Family population histograms
  restricted to the transcription factor functional category (black
  circles are simulations, magenta lines empirical data).  (d) Family
  population histograms restricted to the metabolism functional
  category (black circles are simulations, magenta lines empirical
  data).  }
  \label{fig:supp-comparision}
\end{figure}

\begin{figure}[tbp]
\begin{center}
\includegraphics[width=0.9\columnwidth]{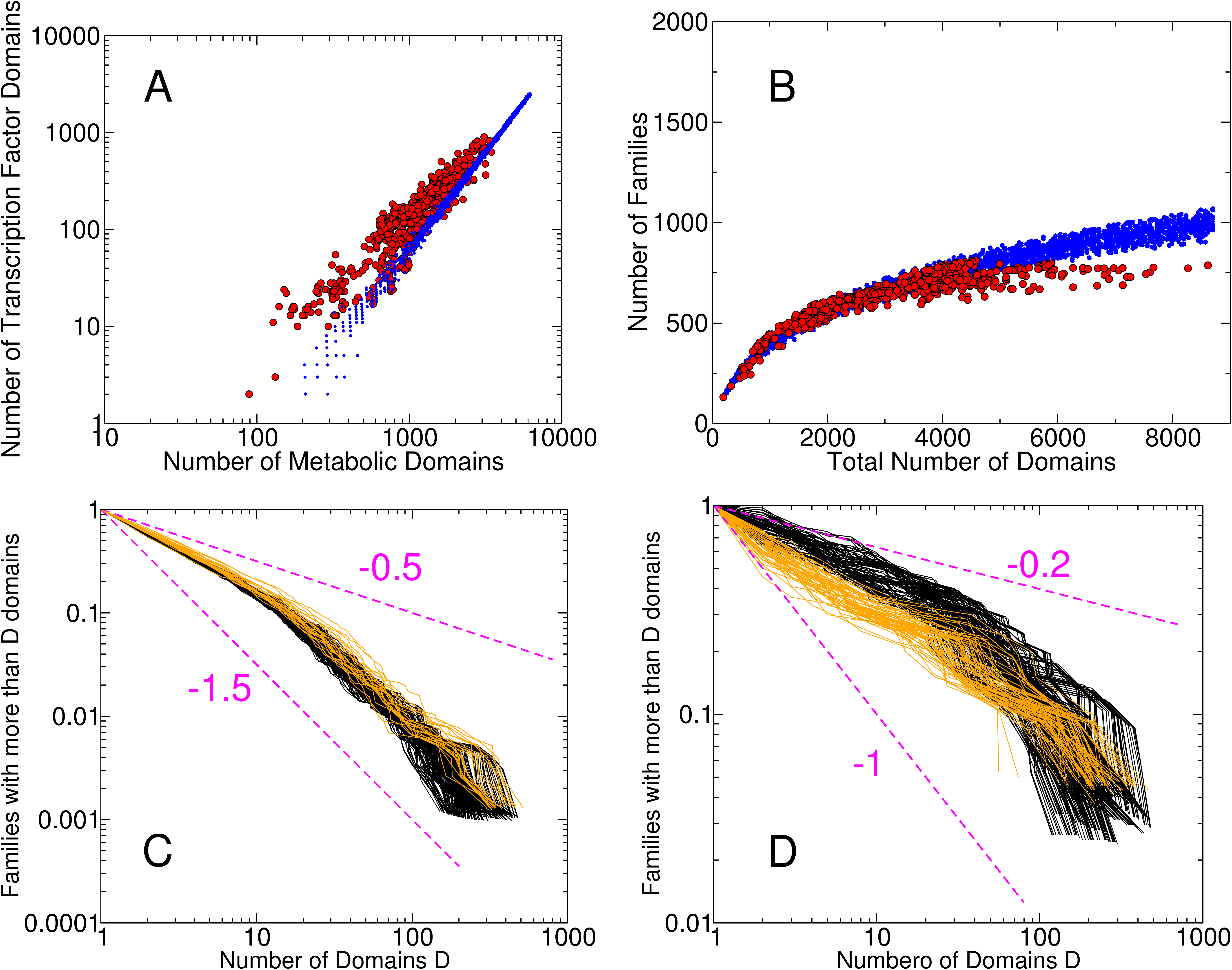}
\end{center}
\caption{ \rev{ { \bf Simulations of the correlated duplication model
      Ia for two categories (transcription factors and metabolic
      enzymes).} The plots are obtained from $1000$ realizations with
    $\alpha=0.3$, $\theta=140$ and $U=7000$. The observables are the
    same as in figure \ref{fig:supp-comparision}. (a) scaling of the
    number of transcription factors with the number of metabolic
    enzymes. (b) Number of families as a function of genome size
    $n$. (c) Family population (cumulative) histograms. (c) Family
    population histograms restricted to the families belonging to the
    transcription factor functional category. } }
  \label{fig:supp-ToolPan}
\end{figure}

\begin{figure}[tbp]
\begin{center}
\includegraphics[width=0.9\columnwidth]{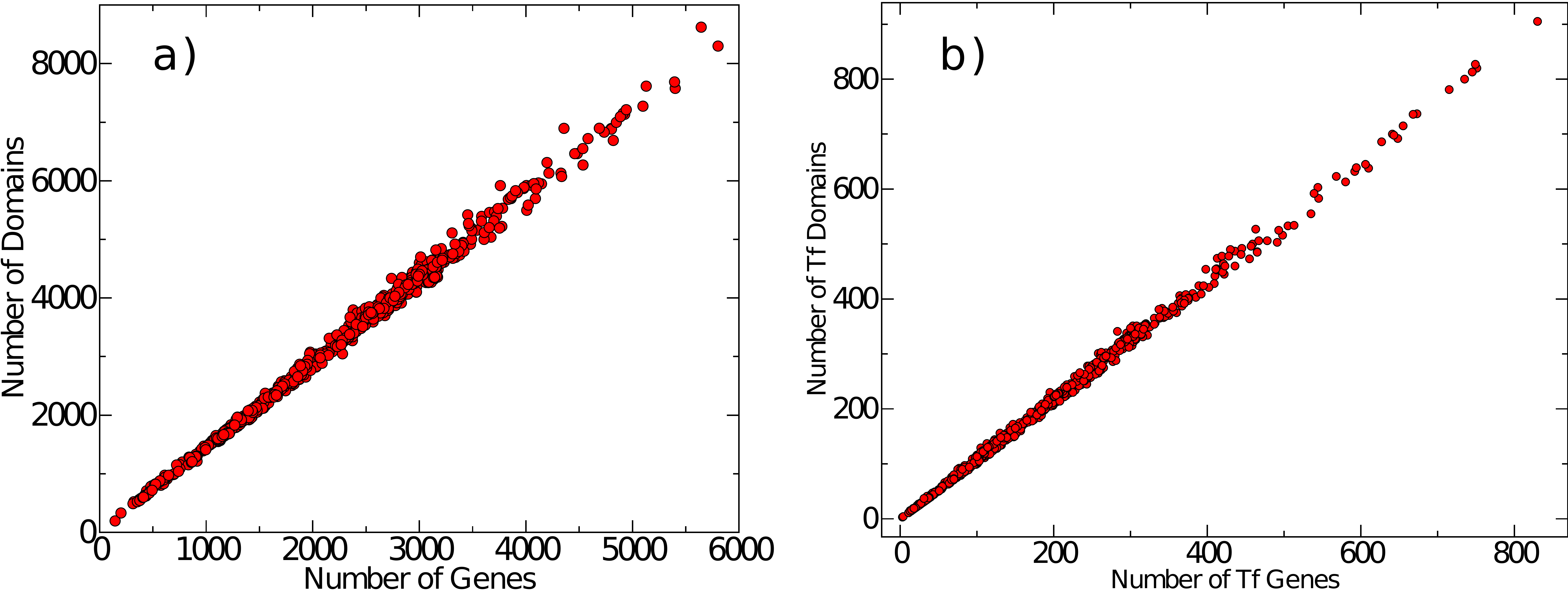}
\end{center}
\caption{ { \bf Linear relation between the number of domains and the
    number of genes.}  (a) Number of Domains vs. number of
  protein-coding genes for the 753 bacteria in the SUPERFAMILY
  database. There are, on average, $ 1.45 $ domains per gene. (b)
  Linear scaling behaviour of the number of TF domains vs.  number of
  TF genes. There are, on average, $ 1.09 $ TF domains in a TF gene.}
  \label{fig:supp-gene}
\end{figure}

\begin{figure}[tbp]
\begin{center}
\includegraphics[width=0.6\columnwidth]{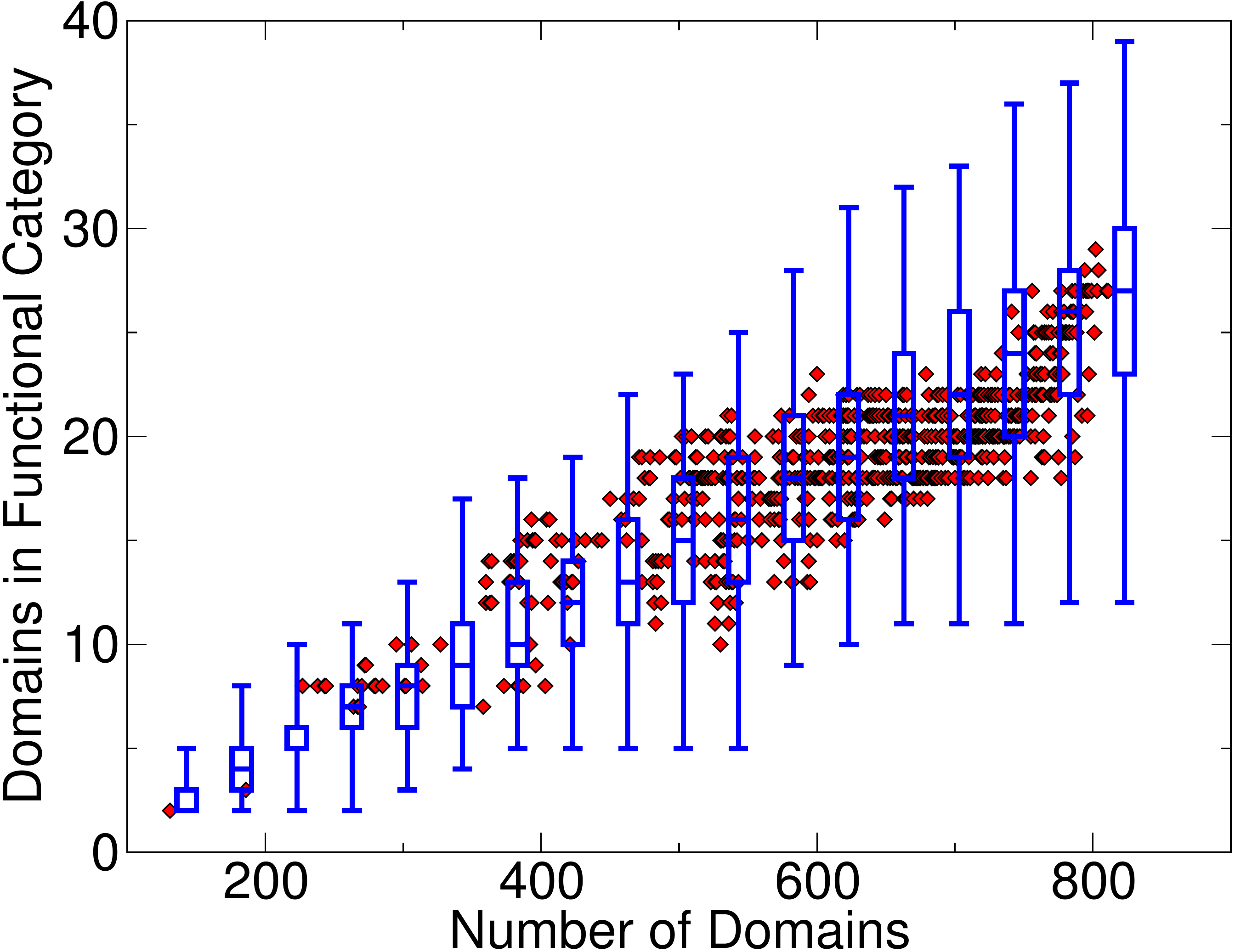}
\end{center}
\caption{ { \bf Simulation of the number of transcription factor
    families.}  Comparison between empirical data and simulations of
  the number of transcription factor domain families plotted against
  total number of families.  The scaling is empirically linear,
  i.e. the number of TF domain families is reproduced by a null
  hypothesis of independent addition model. The choice of the
  parameter is $ 0.035 $. }
  \label{fig:supp-simTFsupfam}
\end{figure}

\begin{figure}[tbp]
\begin{center}
\includegraphics[width=0.9\columnwidth]{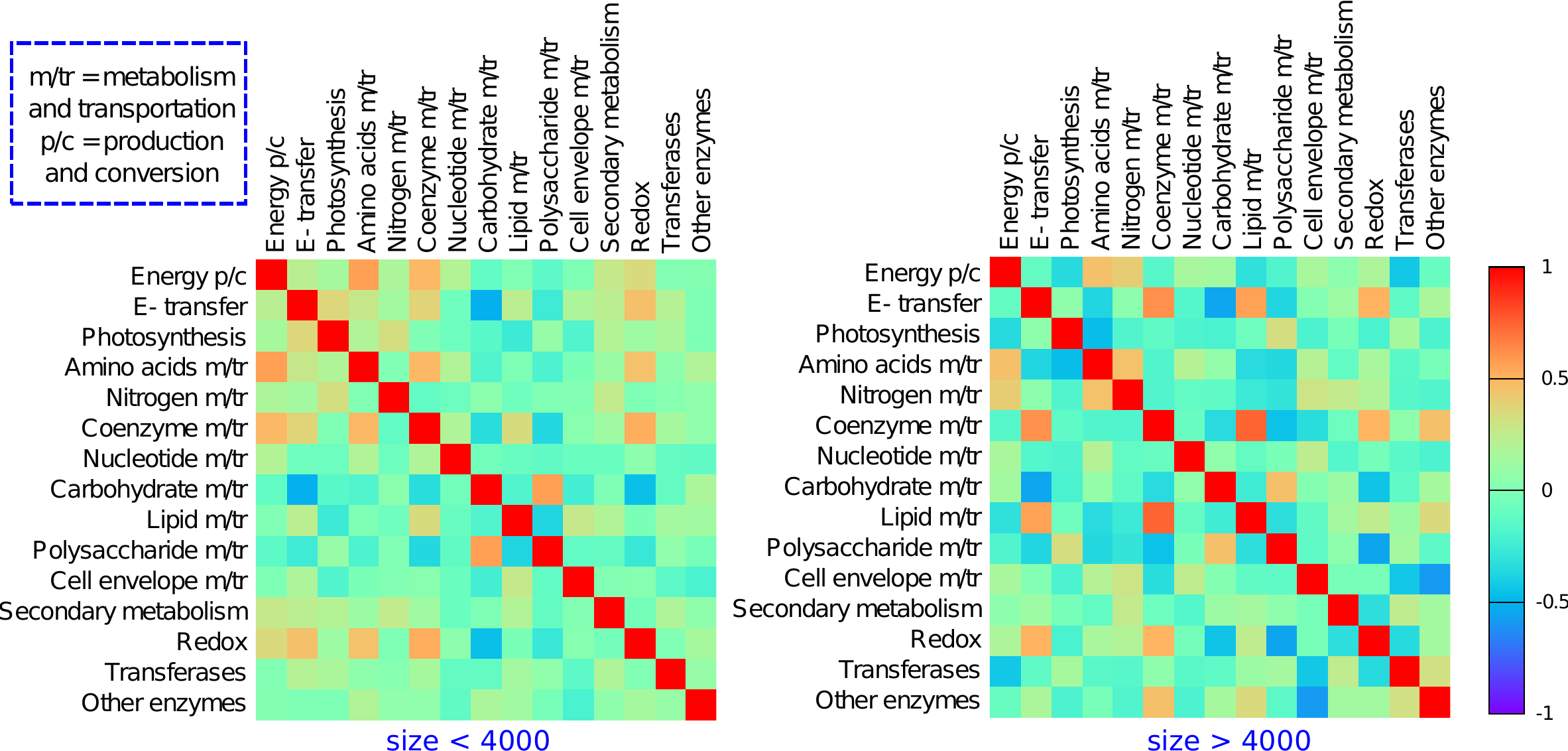}
\end{center}
\caption{ { \bf Correlation matrix for two sets of genomes with
    different sizes. } Left panel: Correlation matrix for genomes with
  size $ < 4000 $.  Right panel: Correlation matrix for genome with
  size $ > 4000 $.  The correlations do not depend on size.}
  \label{fig:supp-metcorrsize}
\end{figure}

\begin{figure}[tbp]
\begin{center}
\includegraphics[width=0.9\columnwidth]{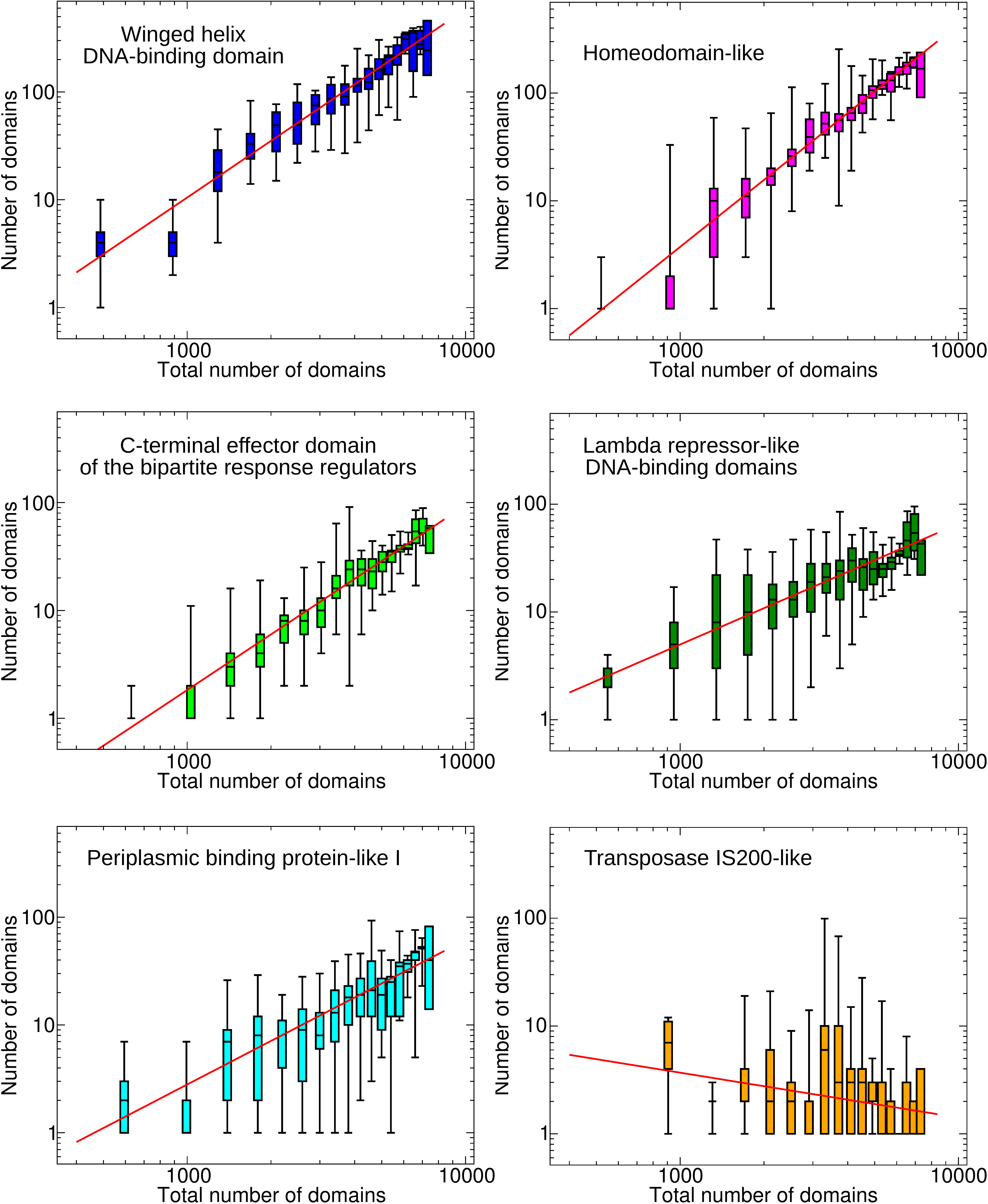}
\end{center}
\caption{ { \bf Most populated transcription factor superfamilies.}
  Boxplots for the population of the six most populated superfamilies
  of TF DNA-binding domains (y-axis in each panel) versus number of
  domains of each genome (x-axis in each panel).  The presence of
  scaling laws appears likely for the three most populated families
  and arguable for the first five. Red lines represent best power law
  fit ($ 1.8 $ for Winged Helix ,$ 2.1 $ for Homeodomain-like and $
  1.7 $ for C-terminal effector)}
  \label{fig:supp-TFclass}
\end{figure}

\begin{figure}[tbp]
\begin{center}
\includegraphics[width=0.6\columnwidth]{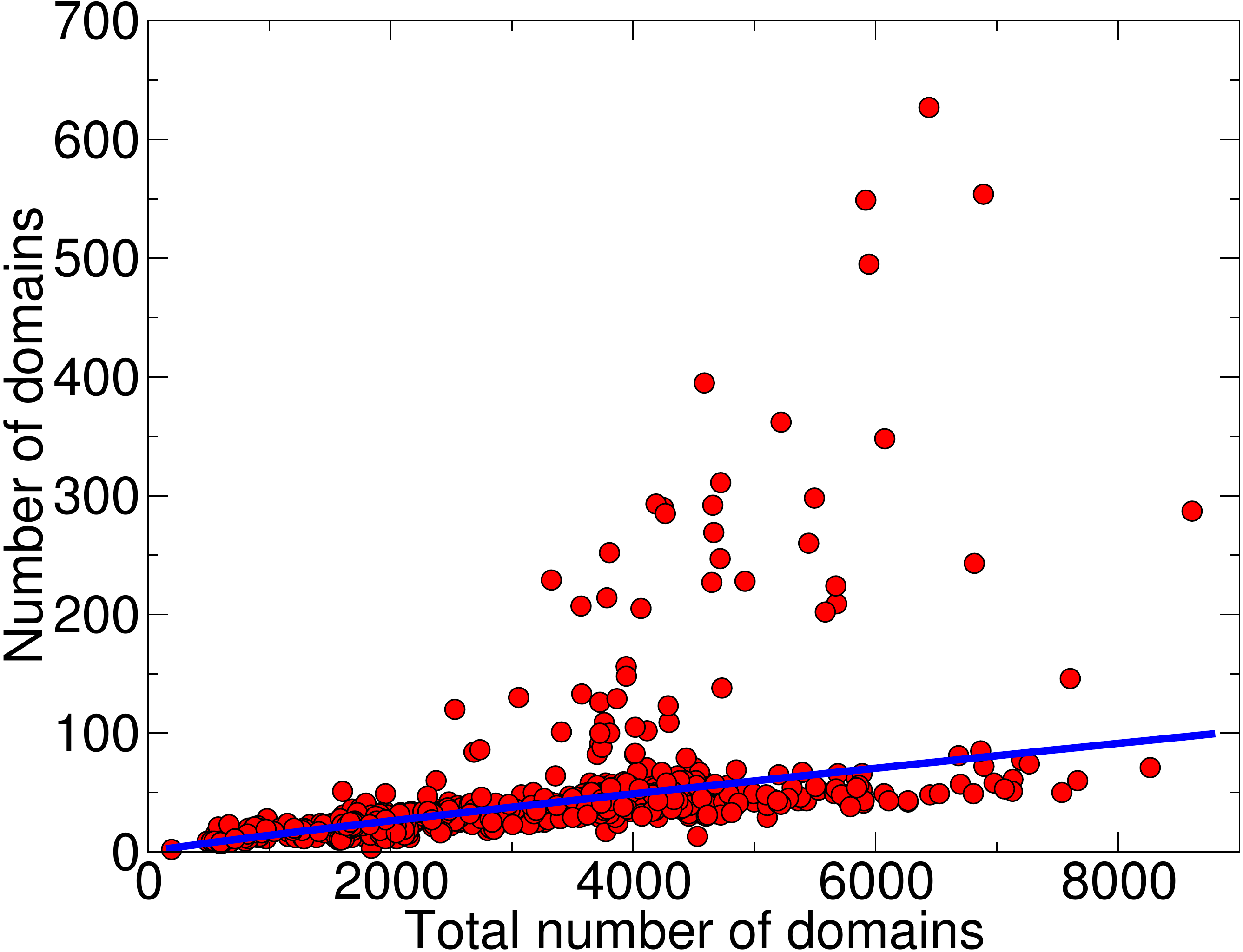}
\end{center}
\caption{\rev{ { \bf Scaling of the least populated transcription factor superfamilies.}
    Collective scaling of the number of
    transcription factor domains after removing the six globally most
    populated families.  While a few genomes show large fluctuations
    from the typical trend, a clear scaling is still observable for
    most genomes, with a fitted exponent equal to $0.9$} }
  \label{fig:supp-smalltf}
\end{figure}

\begin{figure}[tbp]
\begin{center}
\includegraphics[width=0.6\columnwidth]{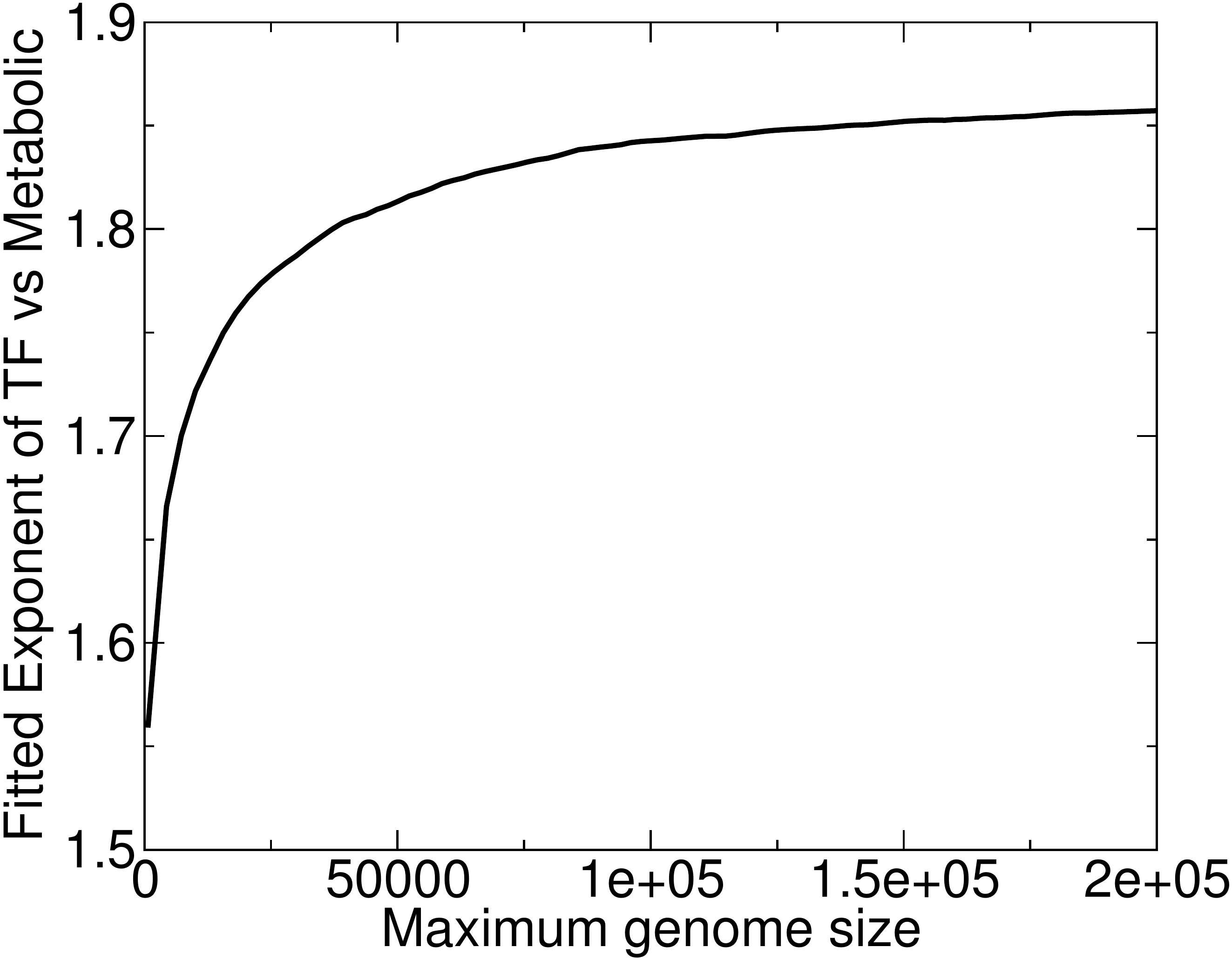}
\end{center}
\caption{MARCO\rev{ { \bf Finite-size effects on the scaling exponent
      $\zeta_{TF}$ for transcription factors in simulations of model
      Ib.}  The plot shows the fitted exponent (y-axis) from the curve
    of the number of transcription factor domains versus the number of
    metabolic enzymes in $500$ simulated realizations of model Ib with
    parameter $\alpha=0.3$ and $\theta=140$. Each point on the x-axis
    corresponds to simulated data stopped at a given size $n$. The
    mean-field prediction ($\zeta_{TF} = 2$) is reached only in the
    limit $n \to \infty$. This plot shows that the fitted exponent
    $1.6$ (instead of $2$) for the growth of transcription factors vs
    metabolic domains is due to a finite-size effect of a process that
    produces an exponent $2$ in the large-$n$ limit. The same effect
    is present in models Ia and II.} }
  \label{fig:supp-ExpSim}
\end{figure}

\begin{figure}[tbp]
\begin{center}
\includegraphics[width=0.6\columnwidth]{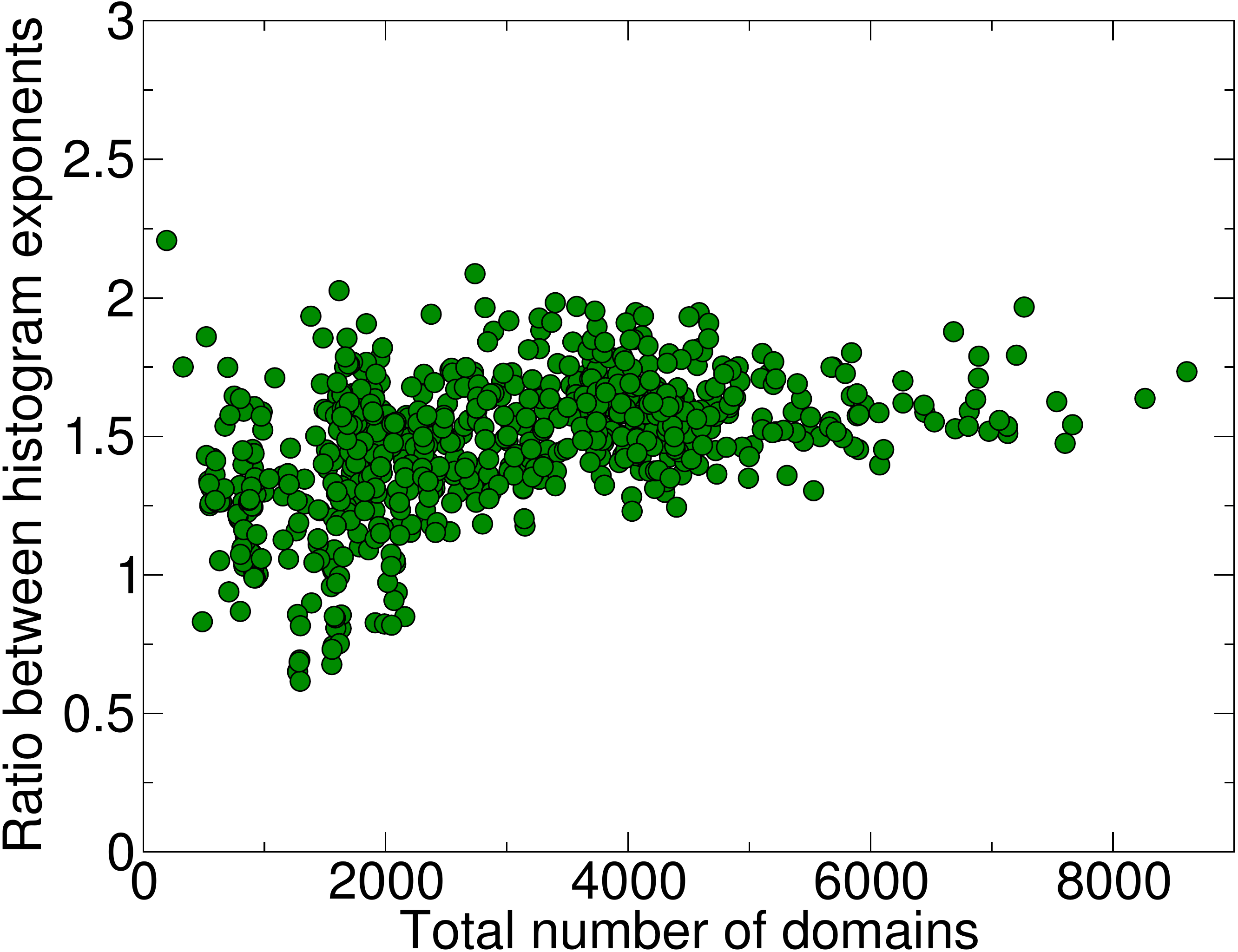}
\end{center}
\caption{\rev{ { \bf Ratio between exponents of family population
      histograms.} The plot reports the ratio $\beta / \beta_{TF}$
    between the exponent of the total family population histograms and
    the histograms restricted to the transcription factor families
    (see Figure~\ref{fig:histoexp} in the main text), as a function of
    genome size.  The values of the ratio are distributed around $1.6$
    and the fluctuation range decreases with increasing genome size.}
}
  \label{fig:supp-scatterExp}
\end{figure}

\clearpage

\newpage



\begin{table}[tbp]
  \caption{ {\bf   
      Fitted values of $\chi_c$ and offsets $A_c$ from  $f_c$ vs $f$ for the ten largest functional  categories} }
 \begin {tabular}{|c|c|c|c|}
   \hline
     & $A_c$  & $ \chi_c $ & Reduced chi square \\
   \hline
   Transcription Factors     & $ 2.2 \pm 0.4 $  & $ 0.0267 \pm 0.0006 $ & $4.5$ \\
   \hline
   Translation  & $ 61.0 \pm 0.35 $ &  $ 0.0133 \pm 0.0006 $ & $3.9$\\
   \hline
   Small molecule binding             & $ 3.0 \pm 0.2 $ & $ 0.01 \pm 0.0002 $  & $0.9$\\
   \hline
   Nucleotide transport and metabolism     & $ 5.6 \pm 0.3 $  & $ 0.02 \pm 0.0005 $ & $3.1$ \\
   \hline
   DNA replication/repair  & $ 9.5 \pm 0.6 $ &  $ 0.0437 \pm 0.0009 $ & $9.8$\\
   \hline
   Inorganic ion transport and metabolism           & $ 0.2 \pm 0.4 $ & $ 0.0272 \pm 0.0005 $  & $3.5$\\
   \hline
   Redox     & $ -7.6 \pm 0.5 $  & $ 0.0592 \pm 0.0008 $ & $7.9$ \\
   \hline
   Transferases  & $ 5.3 \pm 0.2 $ &  $ 0.0213 \pm 0.0004 $ & $1.6$\\
   \hline
   Other enzymes            & $ -14.8 \pm 1.1 $ & $ 0.155 \pm 0.002 $  & $35.7$\\
   \hline
   Signal transduction            & $ -3.2 \pm 0.3 $ & $ 0.0282 \pm 0.0005 $  & $3.3$\\
   \hline
 \end{tabular}
 \begin{flushleft} 
 \rev{   The number of evolutionary families belonging to a functional
   category follows a linear law in empirical data. The table reports
   fits of $f_c = A_c + \chi_c f$ from the plots in
   Figure~\ref{fig:Chi_c} of the main text, where $f_c$ represents the
   number of families in category $c$ on all genomes and $f$ is the
   total number of families on the genome.  The third column is the
   reduced chi square. }
 \end{flushleft}
 \label{tab:fit-catfam}
 \end{table}

\begin{table}[tbp]
 \caption{ {\bf \rev{Data of fitted exponents from Figure \ref{fig:expcorr}
     of the main text, for the ten largest functional categories}}  }
 \begin {tabular}{|c|c|c|}
   \hline
     & $\zeta_c$  & $ \beta_c $  \\
   \hline
   Transcription Factors     & $ 1.6 \pm 0.02 $  &  $ 0.47 \pm 0.01  $  \\
   \hline
   Translation  & $ 0.176 \pm 0.003   $ &  $ 1.46 \pm 0.02 $ \\
   \hline
   Small molecule binding     & $ 0.918 \pm 0.006  $ & $  0.25 \pm 0.01  $ \\
   \hline
   Nucleotide transport and metabolism     & $  0.61 \pm 0.01 $  & $
   0.71 \pm 0.01  $  \\ 
   \hline
   DNA replication/repair  & $  0.54 \pm 0.01  $ &  $  0.9 \pm 0.01 $ \\
   \hline
   Inorganic ion transport and metabolism   & $  1.40 \pm 0.02  $ & $
   0.46 \pm 0.01  $  \\ 
   \hline
   Redox     & $  1.3 \pm 0.01  $  & $  0.52 \pm 0.02  $  \\
   \hline
   Transferases  & $  1.09 \pm 0.01  $ &  $  0.43 \pm 0.01  $ \\
   \hline
   Other enzymes   & $  1.09 \pm 0.01  $ & $  0.64 \pm 0.01  $ \\
   \hline
   Signal transduction  & $  1.77 \pm 0.03  $ & $ 0.4 \pm 0.01  $ \\
   \hline
 \end{tabular}
 \begin{flushleft} 
 \end{flushleft}
 \label{tab:fit-LAW}
 \end{table}

\begin{table}[!ht]
\caption{
\bf{Correlation coefficients between the populations of metabolic
	functional categories}}
\resizebox{\columnwidth}{!}{
\begin{tabular}{|c|c|c|c|c|c|c|c|c|c|c|c|c|c|c|c|}
\hline
& En &  e- & Ph & Aa  & N & Co & Nu & Ca & Li & Ps & Ce & 2M & Rx & Tr & Ot \\
\hline
En & $ 1 $ & $ 0.14 $ & $ 0.07$ & $ 0.55 $ & $ 0.23 $ & $ 0.36 $ & $ 0.19 $ &
 $ -0.06 $ & $ -0.08 $ & $ -0.14 $ & $ 0.02 $ & $ 0.22 $ & $ 0.31 $ & $ -0.10 $ & $ -0.004 $ \\
\hline
 e- & $ 0.14 $ & $ 1 $ & $ 0.29  $ & $ 0.15 $ & $ 0.11 $ & $ 0.43 $ & $ -0.09 $ &
 $ -0.52 $ & $ 0.35 $ & $ -0.29 $ & $ 0.13 $ & $ 0.19 $ & $ 0.47 $ & $ 0.09 $ & $ 0.05 $ \\
\hline
Ph & $ 0.07 $ & $ 0.29 $ & $ 1  $ & $ 0.12 $ & $ 0.21 $ & $ -0.02 $ & $ -0.09 $ &
$ -0.16 $ & $ -0.21 $ & $ 0.14 $ & $ -0.18 $ & $ 0.15 $ & $ 0.06 $ & $ 0.16 $ & $ -0.05 $ \\
\hline
Aa  & $ 0.55 $ & $ 0.15 $ & $ 0.12  $ & $ 1 $ & $ 0.08 $ & $ 0.39 $ & $ 0.19 $ &
$ -0.14 $ & $ -0.07 $ & $ -0.22 $ & $ 0.01 $ & $ 0.07 $ & $ 0.40 $ & $ 0.02 $ & $ 0.14 $ \\
\hline
N & $ 0.23 $ & $ 0.11 $ & $ 0.21  $ & $ 0.08 $ & $ 1 $ & $ -0.13 $ & $ -0.08 $ &
 $ -0.003 $ & $ -0.14$ & $ -0.09 $ & $ 0.09 $ & $ 0.26 $ & $ 0.04 $ & $ -0.03 $ & $ -0.02 $ \\
\hline
Co & $ 0.36 $ & $ 0.43 $ & $ -0.02  $ & $ 0.39 $ & $ -0.13$ & $ 1 $ & $ 0.14 $ &
 $ -0.33 $ & $ 0.44 $ & $ -0.37 $ & $ -0.04 $ & $ 0.08 $ & $ 0.51 $ & $ 0.12 $ & $ 0.16 $ \\
\hline
Nu & $ 0.19 $ & $ -0.09 $ & $ -0.09  $ & $ 0.19 $ & $ -0.08 $ & $ 0.14 $ & $ 1 $
 & $ -0.03 $ & $ -0.09 $ & $ -0.10 $ & $ -0.02 $ & $ -0.10 $ & $ 0.03 $ & $ -0.11 $ & $ -0.13 $ \\
\hline
Ca & $ -0.06 $ & $ -0.52 $ & $ -0.16  $ & $ -0.14 $ & $ -0.003 $ & $ -0.33 $ &
$ -0.03 $ & $ 1 $ & $ -0.20 $ & $ 0.53 $ & $ -0.18 $ & $ 0.02 $ & $ -0.46 $ & $ -0.11 $ & $ 0.16 $ \\
\hline
Li & $ -0.08 $ & $ 0.35 $ & $ -0.21  $ & $ -0.07 $ & $ -0.14 $ & $ 0.44 $ &
$ -0.09 $ & $ -0.20 $ & $ 1 $ & $ -0.35 $ & $ 0.15 $ & $ 0.18 $ & $ 0.06 $ & $ 0.13 $ & $ 0.20 $ \\
\hline
Ps & $ -0.14 $ & $ -0.29 $ & $ 0.14  $ & $ -0.22 $ & $ -0.09 $ & $ -0.37 $ &
 $ -0.10 $ & $ 0.53 $ & $ -0.35 $ & $ 1 $ & $ -0.12 $ & $ -0.05 $ & $ -0.36 $ & $ 0.09 $ & $ -0.07 $ \\
\hline
Ce & $ 0.02 $ & $ 0.13 $ & $ -0.18  $ & $ 0.01 $ & $ 0.09 $ & $ -0.04 $ & $ -0.02 $ &
 $ -0.18 $ & $ 0.15 $ & $ -0.12 $ & $ 1 $ & $ -0.0002 $ & $ 0.01 $ & $ -0.22 $ & $ -0.31 $ \\
\hline
2M & $ 0.22 $ & $ 0.19 $ & $ 0.15  $ & $ 0.07 $ & $ 0.26 $ & $ 0.08 $ & $ -0.10 $ &
 $ 0.02 $ & $ 0.18 $ & $ -0.05 $ & $ -0.0002 $ & $ 1 $ & $ -0.11 $ & $ 0.20 $ & $ 0.08 $ \\
\hline
Rx & $ 0.31 $ & $ 0.47 $ & $ 0.06  $ & $ 0.40 $ & $ 0.04 $ & $ 0.51 $ & $ 0.03 $ & $ -0.46 $ &
 $ 0.06 $ & $ -0.36 $ & $ 0.01 $ & $ -0.11 $ & $ 1 $ & $ -0.10 $ & $ 0.14 $ \\
\hline
Tr & $ -0.10 $ & $ 0.09 $ & $ 0.16  $ & $ 0.02 $ & $ -0.03 $ & $ 0.12 $ & $ -0.11 $ &
$ -0.11 $ & $ 0.13 $ & $ 0.09 $ & $ -0.22 $ & $ 0.20 $ & $ -0.10 $ & $ 1 $ & $ 0.17 $ \\
\hline
Ot & $ -0.004 $ & $ 0.05 $ & $ -0.05  $ & $ 0.14 $ & $ -0.02 $ & $ 0.16 $ & $ -0.13 $
 & $ 0.16 $ & $ 0.20 $ & $ -0.07 $ & $ -0.31 $ & $ 0.08 $ & $ 0.14 $ & $ 0.17 $ & $ 1 $ \\
\hline
\end{tabular}
}
\begin{flushleft}
  Pearson's correlation coefficients between the populations of 24
  different metabolic functional categories from the SUPERFAMILY
  database for 753 bacteria. Correlations are calculated from
  fluctuations of categories from the average trend (see
  Methods). Both correlation and anticorrelation are present between
  categories. Metabolism categories are highly (anti-)correlated.  We
  used the following short forms for the metabolic functional
  categories: En = Energy p/c, e- = Electrons transfer, Ph =
  Photosynthesis, Aa = Amino acids m/tr, N = Nitrogen m/tr, Co =
  Coenzyme m/tr, Nu = Nucleotide m/tr, Ca = Carbohydrate m/tr, Li =
  Lipid m/tr, Ps = Polysaccharide m/tr, Ce = Cell envelope m/tr, 2M =
  Secondary metabolism, Rx = Redox, Tr = Transferases, Ot = Other
  enzymes.  Where m/tr stands for ``metabolism and trasportation'' and
  p/c means ``production and conversion''.
\end{flushleft}
\label{tab:supp-correlation}
\end{table}

\begin{table}[!ht]
\caption{
\bf{P-Values of correlation coefficients between the populations of metabolic
	functional categories}}
\resizebox{\columnwidth}{!}{
\begin{tabular}{|c|c|c|c|c|c|c|c|c|c|c|c|c|c|c|c|}
\hline
& En &  e- & Ph & Aa & N & Co & Nu & Ca & Li & Ps & Ce & 2M & Rx & Tr & Ot \\
\hline
En &$ 0 $ &  $ \mathbf {5 \cdot 10^{-5}} $ & $ \mathbf { 0.02 }  $ & $ \mathbf { <10^{-6}} $ & $ \mathbf { <10^{-6}} $ &
$\mathbf { <10^{-6}} $ &$\mathbf { <10^{-6}} $ & $ 0.05 $ & $\mathbf { 0.01} $ & $\mathbf { 4 \cdot 10^{-5} }$ & $ 0.26 $ &
$ \mathbf {<10^{-6} }$ & $ \mathbf {<10^{-6}} $ & $ \mathbf {4 \cdot 10^{-3}} $ & $ 0.46 $ \\
\hline
e- &$ \mathbf {5 \cdot 10^{-5} }$ & $ 0 $ & $\mathbf { <10^{-6}  }$ & $ \mathbf { 2 \cdot 10^{-5}} $ &$ \mathbf {1 \cdot 10^{-3}} $ &
$\mathbf { <10^{-6} }$ & $\mathbf { 8 \cdot 10^{-3} }$ & $\mathbf {  <10^{-6}} $ & $ \mathbf { <10^{-6}} $ &
$  \mathbf {<10^{-6} }$ & $ \mathbf {3 \cdot 10^{-4}} $ & $ \mathbf {1 \cdot 10^{-6} }$ & $\mathbf { <10^{-6} }$ & $\mathbf {7 \cdot 10^{-3}}$ & $ 0.08 $ \\
\hline
Ph &$ \mathbf {0.02} $ & $ \mathbf {<10^{-6}} $ & $ 0  $ & $ \mathbf {1 \cdot 10^{-3}} $ & $ \mathbf {<10^{-6}} $ & $ 0.29 $ &
$ \mathbf {2 \cdot  10^{-3}} $ & $\mathbf { <10^{-6}} $ & $ \mathbf {<10^{-6} }$ & $\mathbf { 2 \cdot 10^{-4}} $ & $ \mathbf {<10^{-6} }$
& $ \mathbf {5 \cdot 10^{-5}} $ & $ 0.06 $ & $ \mathbf {2 \cdot 10^{-5}} $ & $ 0.08 $ \\
\hline
Aa & $ \mathbf {<10^{-6}} $ & $ \mathbf {2 \cdot 10^{-5}} $ & $ \mathbf {1 \cdot 10^{-3}}  $ & $ 0 $ & $ \mathbf {0.02} $ &
$ \mathbf {<10^{-6}} $ & $ \mathbf {2 \cdot 10^{-6}} $ & $ \mathbf {4 \cdot 10^{-5}} $ & $ \mathbf {0.02} $ & $ \mathbf {<10^{-6}} $ &
$ 0.39 $ & $\mathbf { 0.03} $ & $\mathbf { <10^{-6}} $ & $ 0.28 $ & $ \mathbf {5 \cdot 10^{-5}} $ \\
\hline
N &$ \mathbf { <10^{-6}} $ & $ \mathbf {1 \cdot 10^{-3}} $ & $\mathbf { <10^{-6}}  $ & $ \mathbf {0.02} $ & $ 0 $ & $\mathbf { 2 \cdot 10^{-4}} $
& $ \mathbf {0.01} $ & $ 0.47 $ & $ \mathbf {7 \cdot 10^{-5} }$ & $ \mathbf {5 \cdot 10^{-3}} $ & $ \mathbf {8 \cdot 10^{-3}} $ &
$ \mathbf {<10^{-6}} $ & $ 0.13 $ & $ 0.18 $ & $ 0.31 $ \\
\hline
Co &$ \mathbf {<10^{-6} }$ & $\mathbf { <10^{-6} }$ & $ 0.29  $ & $ \mathbf {<10^{-6}} $ & $ \mathbf {2 \cdot 10^{-4}} $ & $ 0 $ &
$\mathbf { 1 \cdot 10^{-4}} $ & $\mathbf { <10^{-6} }$ & $ \mathbf {<10^{-6}} $ & $ \mathbf {<10^{-6}} $ & $ 0.13 $ & $ \mathbf {0.02} $ &
$ \mathbf {<10^{-6}} $ & $ \mathbf {2 \cdot 10^{-4}} $ & $\mathbf { 4 \cdot 10^{-6}} $ \\
\hline
Nu & $\mathbf { <10^{-6}} $ & $  \mathbf {8 \cdot 10^{-3}} $ & $ \mathbf {2 \cdot 10^{-3}}  $ & $ \mathbf {2 \cdot 10^{-6}} $
& $\mathbf { 0.01} $ & $ \mathbf {1 \cdot 10^{-4}} $ & $ 0 $ & $ 0.20 $ & $ \mathbf {5 \cdot 10^{-3}} $ & $ \mathbf {3 \cdot 10^{-3}} $
& $ 0.26 $ & $ \mathbf {3 \cdot 10^{-3}} $ & $ 0.17 $ & $\mathbf { 8 \cdot 10^{-3}} $ & $ \mathbf {9 \cdot 10^{-5}} $ \\
\hline
Ca & $ 0.05 $ & $ \mathbf { <10^{-6}} $ & $\mathbf { <10^{-6} } $ & $\mathbf { 4 \cdot 10^{-5}} $ & $ 0.47 $ &
$ \mathbf {<10^{-6}} $ & $ 0.20 $ & $ 0 $ & $\mathbf { <10^{-6}} $ & $ \mathbf {<10^{-6} }$ & $\mathbf { <10^{-6}} $ & $ 0.30 $ &
$ \mathbf {<10^{-6}} $ & $ \mathbf {8 \cdot 10^{-4}} $ & $\mathbf { 7 \cdot 10^{-6}} $ \\
\hline
Li & $\mathbf { 0.01} $ & $ \mathbf {<10^{-6}}  $ & $ \mathbf {<10^{-6}} $ & $ \mathbf {0.02} $ & $ \mathbf {7 \cdot 10^{-5}} $ &
$ \mathbf {<10^{-6}} $ &
$\mathbf { 5 \cdot 10^{-3}} $ & $ \mathbf {<10^{-6}} $ & $ 0 $ & $\mathbf { <10^{-6}} $ & $ \mathbf {3 \cdot 10^{-5}} $ & $\mathbf { <10^{-6}} $ &
$ 0.06 $ & $ \mathbf {3 \cdot 10^{-4}} $ & $\mathbf { 2 \cdot 10^{-6}} $ \\
\hline
Ps & $\mathbf { 4 \cdot 10^{-5}} $ & $ \mathbf {<10^{-6} }$ & $\mathbf { 2 \cdot 10^{-4}}  $ & $ \mathbf {<10^{-6}} $ &
$\mathbf { 5 \cdot 10^{-3}} $ & $\mathbf { <10^{-6} }$ & $\mathbf { 3 \cdot 10^{-3}} $ & $ \mathbf {<10^{-6}}$ & $\mathbf { <10^{-6}} $ & $ 0 $ &
$\mathbf { 5 \cdot 10^{-4}} $ & $ 0.07 $ & $ \mathbf {<10^{-6}} $ & $\mathbf { 6 \cdot 10^{-3}} $ & $\mathbf { 0.03 }$ \\
\hline
Ce &$ 0.26 $ & $ 3 \cdot 10^{-4}  $ & $ <10^{-6}  $ & $ 0.39 $ & $ 8 \cdot 10^{-3} $ &
$ 0.13 $ & $ 0.26 $ & $ <10^{-6} $ & $ 3 \cdot 10^{-5} $ & $ 5 \cdot 10^{-4} $ & $ 0 $ &
$ 0.50 $ & $ 0.38 $ & $ \mathbf{<10^{-6}} $ & $ \mathbf{<10^{-6}} $ \\
\hline
2M & $ \mathbf {<10^{-6}} $ & $ \mathbf {1 \cdot 10^{-6} } $ & $ \mathbf {5 \cdot 10^{-5}}  $ & $\mathbf { 0.03 }$ & $\mathbf { <10^{-6}} $ &
$ \mathbf {0.02} $ & $ \mathbf {3 \cdot 10^{-3}} $ & $ 0.30 $ & $\mathbf { <10^{-6}} $ & $ 0.07 $ & $ 0.50 $ & $ 0 $ &
$\mathbf { 8 \cdot 10^{-4} }$ & $\mathbf { <10^{-6}} $ & $ \mathbf {0.01} $ \\
\hline
Rx & $ \mathbf {<10^{-6}} $ & $\mathbf { <10^{-6}} $ & $ 0.06  $ & $\mathbf { <10^{-6}} $ & $ 0.13 $ & $ \mathbf {<10^{-6}} $ & $ 0.17 $ &
$ \mathbf {<10^{-6}} $ & $ 0.06 $ & $\mathbf { <10^{-6} }$ & $ 0.38 $ & $\mathbf { 8 \cdot 10^{-4}} $ & $ 0 $ &
$ \mathbf {3 \cdot 10^{-3}} $ & $ \mathbf {4 \cdot 10^{-5} }$ \\
\hline
Tr & $ \mathbf {4 \cdot 10^{-3}} $ & $ \mathbf {7 \cdot 10^{-3}} $ & $ \mathbf {2 \cdot 10^{-5}}  $ & $ 0.28 $ & $ 0.18 $ &
$\mathbf { 2 \cdot 10^{-4}} $ & $\mathbf { 8 \cdot 10^{-4}} $ & $ \mathbf {8 \cdot 10^{-4}} $ & $ \mathbf {3 \cdot 10^{-4}} $ &
$\mathbf { 6 \cdot 10^{-3}} $ & $ \mathbf {<10^{-6} }$ & $\mathbf { <10^{-6}} $ & $ \mathbf {3 \cdot 10^{-3}} $ & $ 0 $ & $ \mathbf {1 \cdot 10^{-6}} $ \\
\hline
Ot & $ 0.46 $ & $ 0.08 $ & $ 0.08  $ & $ \mathbf {5 \cdot 10^{-5}} $ & $ 0.31 $ & $ \mathbf {4 \cdot 10^{-6}} $ &
$ \mathbf {9 \cdot 10^{-5}} $ & $ \mathbf {7 \cdot 10^{-6}} $ & $ \mathbf {2 \cdot 10^{-6}} $ & $ \mathbf {0.03} $ & $ \mathbf {<10^{-6}} $ &
$ \mathbf {0.01} $ & $ \mathbf {4 \cdot 10^{-5}} $ & $ \mathbf {1 \cdot 10^{-6}} $ & $ 0 $ \\
\hline
\end{tabular}
}
\begin{flushleft}
  P-values of the Pearson's correlation coefficients between the
  populations of 24 different metabolic functional categories from the
  SUPERFAMILY database for 753 bacteria \rev{(the most significant
    values are in boldface)}. Correlations are calculated from
  fluctuations of categories from the average trend (see Methods).
  The (anti-)correlation is statistically significant for the most of
  the metabolic categories.  We used the following short forms for the
  metabolic functional categories: En = Energy p/c, e- = Electrons
  transfer, Ph = Photosynthesis, Aa = Amino acids m/tr, N = Nitrogen
  m/tr, Co = Coenzyme m/tr, Nu = Nucleotide m/tr, Ca = Carbohydrate
  m/tr, Li = Lipid m/tr, Ps = Polysaccharide m/tr, Ce = Cell envelope
  m/tr, 2M = Secondary metabolism, Rx = Redox, Tr = Transferases, Ot =
  Other enzymes.  Where m/tr stands for ``metabolism and
  trasportation'' and p/c means ``production and conversion''.
\end{flushleft}
\label{tab:supp-pvalcorrelation}
\end{table}

\end{document}